%% file: main.tex
\documentclass[a4paper,11pt]{article}
\usepackage[utf8]{inputenc} 
\usepackage[spanish,english]{babel} 
\usepackage{amsmath,amsfonts,amssymb}
\usepackage{dcolumn}
\usepackage{booktabs}
\usepackage{txfonts} 
\usepackage{authblk} 
\usepackage{graphicx} 
\usepackage{float} 
\usepackage{multirow} 
\usepackage{makecell} 
\usepackage{caption} 
\usepackage{hyperref} 
\usepackage[numbers,comma,sort&compress,square]{natbib}
\makeatletter 
\def\NAT@def@citea{\def\@citea{\NAT@separator}}
\makeatother
\usepackage[table,xcdraw]{xcolor} 
\usepackage[margin=1.5cm]{geometry} 
\usepackage{lineno} 
\usepackage[misc]{ifsym} 
\usepackage{lipsum} 
\usepackage{sidecap} 

\usepackage{comment}
\usepackage{multirow}
\usepackage{colortbl}
\usepackage{subcaption}
\usepackage[capitalize,noabbrev]{cleveref}
 \DeclareMathOperator*{\argmax}{arg\,max}
 \DeclareMathOperator{\sech}{sech}
\usepackage{cancel}
\usepackage[symbols,nogroupskip,nonumberlist,nopostdot=false,abbreviations]
{glossaries-extra}
\usepackage{array}
\usepackage{colortbl}

\usepackage{soul}

\input{glossary}
\usepackage{tcolorbox}
\usepackage{siunitx}

\usepackage{geometry}
\usepackage{media9}
\usepackage{tikz}
\usetikzlibrary{shapes.geometric, arrows, positioning}

\tikzstyle{startstop} = [rectangle, rounded corners, minimum width=3cm, minimum height=1cm, text centered, draw=black, fill=red!30]
\tikzstyle{process}   = [rectangle, minimum width=3cm, minimum height=1cm, text centered, draw=black, fill=blue!20]
\tikzstyle{decision}  = [diamond, aspect=2, text centered, draw=black, fill=green!30]
\tikzstyle{arrow}     = [thick,->,>=stealth]
\usepackage{algorithm}
\usepackage{algpseudocode}
\usepackage{siunitx}
\usepackage{enumerate}
\usepackage{enumitem}
\usepackage{caption}
\usepackage{tcolorbox}

\newcounter{myframecounter}

\crefname{myframecounter}{Box}{Boxes}

\newlist{myenumerate}{enumerate}{1}
\setlist[myenumerate]{label=\textbf{\arabic*.}}

\newlist{myitemize}{itemize}{1}
\setlist[myitemize]{label=\textbf{$\bullet$}}

\newlist{mydescription}{description}{1}
\setlist[mydescription]{font=\normalfont\bfseries}

\usepackage{titlesec}

\titleformat{\subsubsubsection}[runin]
{\normalfont\normalsize\bfseries}{}{0em}{}

\titlespacing*{\subsubsubsection}
{0pt}{3.25ex plus 1ex minus .2ex}{1.5ex plus .2ex}

\newcommand{\subsubsubsection}[1]{\paragraph*{#1}\mbox{}\\}

\makeatletter\renewcommand{\@biblabel}[1]{#1.}\makeatother 
\addto{\captionsenglish}{}
\hypersetup{ 
  colorlinks = true,
  citecolor=  blue,
  linkcolor = {blue},
  filecolor = cyan, 
  urlcolor = blue
}
\DeclareCaptionLabelSeparator{vbar}{~\textbar~}
\DeclareCaptionLabelSeparator{dot}{.~}
\definecolor{lightgray}{gray}{0.9}
\newcommand{\figref}[2][]{%
  \hyperref[{#2}]{%
    Fig.~\ref*{#2}%
    \ifx\\#1\\%
    \else
      #1%
    \fi
  }%
}
\newcommand{\tableref}[1]{%
  \hyperref[{#1}]{%
   Table~\ref*{#1}%
  }%
}
\newcommand{\methodsref}[1]{%
  \hyperref[{methods:#1}]{%
   Methods:~\nameref*{methods:#1}%
  }%
}
\makeatletter 
\renewcommand{\maketitle}{\bgroup\setlength{\parindent}{0pt}
\begin{flushleft}
  \textbf{\LARGE \@title}
  
  \bigskip
  
  \@author
\end{flushleft}\egroup
}
\makeatother
\renewenvironment{abstract} 
 {\par\noindent\textbf{\abstractname~\textbar}\ \ignorespaces}
 {\par\medskip}

\title{Optimal virulence strategies in epidemiological models with asymptomatic transmission}
\author[1,2,3 \Letter]{C. Brandon Ogbunugafor }
\author[2,3]{Sudam Surasinghe}
\affil[1]{Santa Fe Institute, Santa Fe, NM 87501, USA}
\affil[2]{Department of Ecology \& Evolutionary Biology,
Yale University, New Haven, CT 06520, USA}
\affil[3]{Public Health Modeling Unit, Yale School of Public Health, New Haven, CT 06510, USA}

\affil[$\textrm{\Letter}$]{CBO: brandon.ogbunu@yale.edu}

\date{}
  
\begin{document}

\maketitle

\bigskip\bigskip

\begin{abstract}
Asymptomatic infection has gained notoriety as an important feature of infectious disease dynamics. Despite increasing attention, there have been few rigorous examinations of how asymptomatic transmission influences pathogen evolution. In this study, we apply evolutionary invasion analysis to compute optimal strategies for viruses evolving in a system with a distinct asymptomatic transmission stage. We ask how pathogens would evolve under three conditions: with an increase in the mean infectious period in the symptomatic state, with an increase in the mean infectious period in the asymptomatic stage, and an increase in proportion proceeding through the ``mild recovery route" (where the symptomatic state was bypassed entirely). We find that an increased proportion of cases moving through a ``mild recovery route"---which can occur with different host susceptibility or increased public health intervention---leads to a model structure in which mutant pathogens are transmitted largely through the asymptomatic route, with slightly increased evolved virulence levels. In addition, we find that an increase in the mean infectious period of the symptomatic state has a small overall influence on the fitness of the pathogen, when effective transmission can occur via the asymptomatic route. Further, we find that virulence levels change very slightly for both the asymptomatic and symptomatic populations. In sum, our results highlight the evolutionary implications of variation in host susceptibility and public health interventions in the context of asymptomatic transmission. More generally, the findings speak to the need for more nuanced interrogations of subtle routes of transmission, as they can have profound implications in disease evolution, ecology, and epidemiology. 



\end{abstract}

\section{Introduction}
The increased risk of disease emergence has led scientists to focus on previously understudied features of infectious disease dynamics. Modern approaches in disease ecology and epidemiology have provided an appreciation for the heterogeneity that underlies the way pathogens move from host to host\cite{vanderwaal2016heterogeneity}, and include modern concepts such as superspreading \cite{lloyd2005superspreading, althouse2020superspreading} and complex contagion \cite{st2024nonlinear}. Many diseases are defined by multiple transmission mechanisms \cite{kiss2006effect, meszaros_direct_2020}, and appreciation of this complexity has improved the potential to model and predict the dynamics of infectious disease. 


The COVID-19 pandemic presented a special case in which resolving the route of transmission was critical to our attempts to model and predict the dynamics of the outbreak. Specifically, the asymptomatic transmission route is a peculiarity that underlies several challenges to public health practitioners \cite{WeitzAsymptomatic, han2021prevalence, oran2020prevalence}. During the pandemic, asymptomatic individuals were able to transmit SARS-CoV-2 \cite{park2020time, oran2020prevalence, harris2023time}. This facilitated misestimations of the size of outbreaks, incorrect model structures, and insufficient intervention strategies \cite{ WeitzAsymptomatic}.  However, as important as the COVID-19 pandemic was, it is hardly the only infectious disease setting where asymptomatic transmission occurs, as it exists in various viral, bacterial, and parasitic infections \cite{althouse2015asymptomatic, shaikh2023asymptomatic, moghadas2017asymptomatic}. And, more than its biological dimensions, asymptomatic transmission comes with a set of bioethical issues surrounding screening and surveillance \cite{jamrozik2019surveillance, fost1992ethical}. Thus, the larger fields of disease ecology and computational epidemiology should study its general principles and applications more rigorously. 


Although there are not many modeling studies of asymptomatic transmission, even fewer have examined the evolutionary consequences of asymptomatic transmission. Recent work on evolution in the setting of asymptomatic transmission examined evolutionary stable strategies for latency, which controls asymptomatic behavior, and found that small changes in parameters can impact the evolution of the pathogen’s asymptomatic stage, which has substantial implications for disease control strategies\cite{saad2020dynamics}. Further work has highlighted how evolution is affected by host heterogeneity in asympatomatic transmission \cite{saad2021evolution}. The paucity of studies in this arena are notable because this pernicious mode of transmission can influence many aspects of how pathogen populations evolve. The ability of transmission to occur without symptoms or host harm challenges existing frameworks on the constraints on how virulence evolves) \cite{ogbunugafor2024modes}. 

Classical models suggest that there is a relationship between virulence and transmission, and that selection can operate on virulence insofar as it influences transmission \cite{bull_virulence_1994, alizon_virulence_2009,lipsitch_virulence_1997, lenski_evolution_1994}. But if transmission can occur in the absence of symptoms (and associated virulence-related consequences) \cite{park2023intermediate, park2020time}, then traditional models of the evolution of virulence apply less directly. 

In this study, we build on the small but growing literature on the evolutionary implications of natural histories of diseases that contain an asymptomatic route of transmission. Like previous studies in SARS-CoV-2 \cite{saad2020dynamics}, tuberculosis \cite{basu2009evolution}, and HCV \cite{surasinghe2024evolutionary}, we apply evolutionary invasion analysis to mathematical models of infectious diseases, which identify an evolutionary stable strategy (ESS), a concept rooted in evolutionary game theory \cite{smith1973logic}. In infectious disease models, it identifies an optimal uninvadable natural history strategy for a pathogen given a model structure, set of constraints, and choice of decision variable \cite{otto2011biologist}. 

We investigate evolutionary strategies within a multivariate decision-variable framework, focusing on virulence in the context of asymptomatic transmission. We identify ESS strategies for each with respect to how virulence would respond to variation in three properties: (i) an increase in the mean infectious periods of the asymptomatic, (ii) symptomatic stage, and (iii) an increased proportion of cases moving through the ``mild recovery track,'' when hosts bypass the symptomatic class. For each pertinent case, we compare the ESS values for a model containing asymptomatic transmission to a model without one. Importantly, each represents a condition that arises from host heterogeneity in the natural history of the disease, or in the establishment of public health interventions of various kinds.

We find several peculiarities with respect to evolutionary outcomes.  For models with multiple infectious compartments, we find that increases in the proportion proceeding through the mild recovery track select for changes in disease dynamics. In this evolved model, there is a decline in hosts going through the symptomatic transmission route, and asymptomatic transmission becomes predominant with the evolved asymptomatic virulence level slightly higher than in the pre-evolved model (but far below that of the symptomatic case). In addition, in the scenario where the mean infectious period of the symptomatic window increases,  ESS levels can decrease\color{black}. All of these reflect real-world scenarios in which host biology influences the mean infectious period, as in the case of heterogeneity in asymptomatic transmission \cite{saad2021evolution}. 



We summarize our results with respect to modern conversations in mathematical epidemiology and disease evolution. We support recent calls for more nuanced mechanistic studies of infectious disease transmission, as they have consequences for how we intervene in outbreaks and how we consider the evolutionary outcomes of different disease natural histories.

\section{Method}
This article investigates the evolutionary implications of asymptomatic transmission in the context of pathogen evolution. To achieve this objective, we conducted a comparative evolutionary invasion analysis of epidemiological compartmental models with and without the inclusion of asymptomatic host compartments. The analytical framework developed in a previous investigation serves as the basis for this investigation, allowing the computation of \glspl{ess} and facilitating evolutionary invasion analysis for both model structures \cite{surasinghe2024evolutionary}. We focus our inquiry on optimal virulence strategies, as they relate to the adverse clinical outcomes that arise from infectious diseases. 

To provide readers with a clear framework for this study, this section first explains the theoretical principles underlying the evolutionary invasion analysis. Subsequently, we present a detailed outline of the methodological framework similar to that described in previous work \cite{surasinghe2024evolutionary}.  Finally, we present a detailed description of the epidemiological model structure for the host population, which serves as the baseline single-strain pathogen model for this analysis. These models, extended to incorporate resident and mutant strains of the pathogen under the assumption of no superinfection between strains, will be used in the evolutionary invasion analysis presented in the results (\cref{sec:resuls}).

\subsection{Notes and definitions}

Our study focuses, among other things, on how virulence evolves in different model structures.  “Virulence” is a complicated concept used to encapsulate the various negative impacts of pathogen infection on hosts. For our study, we do not equate virulence with host death but rather with any force that takes individuals out of the main model dynamics, no longer able to be infected or infect others. This includes death, as well as symptoms severe enough to remove individuals from the population dynamics modeled in compartmental mathematical frameworks.

Our study is designed to examine the theoretical implications of a pathogen that evolved in different mechanistic transmission models. We use models of viral infectious disease models, and so we will use ``virus" and ``pathogen" interchangeably. We do this to highlight both the specifics of our approach (virus evolution) and emphasize its general utility for examining how asymptomatic transmission may influence pathogen evolution. 

Lastly, our study frequently mentions a ``mild recovery" route, path, or track. This refers to the route where the asymptomatic host population moves directly to the recovered compartment without experiencing symptoms. 

\subsection{Evolutionary invasion analysis}
Evolutionary invasion analysis involves a systematic series of steps to understand how new traits or mutations propagate within a population. Drawing from previous studies on the subject \cite{RN5, williams2021evolutionary, surasinghe2024evolutionary}, the analysis begins with an ecological model that describes the temporal dynamics of the population of interest. In the context of pathogen evolution, these ecological models are defined by epidemiological \glspl{ode}. The focal trait is then identified along with its possible values. In this study, we focus on virulence---the viral characteristics that cause severe illness, rendering an individual bedridden and unable to participate in population circulation---as the trait value while considering the transmission rate as a function of virulence. Additionally, the ecological model is established to capture the frequency of mutations in the trait and the spread of these variants, which may either achieve fixation or face extinction.

Central to this process is the formulation of equations that describe the dynamics of a rare mutant allele introduced into a population dominated by a resident allele. This is often quantified as ``invasion fitness." Because mutations affecting fitness can be rare, the mutant phenotype is analyzed within the ecological context of the resident population at or near its equilibrium. In the analysis of pathogen evolution, stability analysis of a two-strain (resident and mutant) epidemiological model---extending from conventional models---is used to evaluate the invasion fitness and \glspl{ess}. Using linear stability analyses, researchers evaluate whether a mutant type can successfully invade, based on its population growth rate, when introduced as a small perturbation. This step helps to determine the conditions under which a mutant allele can invade the resident population.

Finally, the trait’s evolutionary trajectory is examined with respect to the resident trait value, culminating in identifying resident trait values resistant to invasion by any mutant allele. These trait values are known as \glspl{ess}. This process provides a robust methodology for assessing the dynamics and stability of traits under evolutionary pressure. Based on this process, the general framework for the evolutionary invasion analysis of pathogenesis discussed in \cite{surasinghe2024evolutionary}. 

\subsection{Framework for evolutionary invasion analysis of pathogens}\label{sec:framework}
For increased methodological clarity, we outline the key steps of the framework (proposed by \cite{surasinghe2024evolutionary}) for the evolutionary invasion analysis of pathogens.

\subsubsection*{Step 1: Constructing the epidemiological model and calculating the basic reproduction number}
The first step in the framework for evolutionary invasion analysis involves the construction and analysis of a compartmental model \gls{ode}, such as the Susceptible-Infectious-Recovered (SIR) or Susceptible-Exposed-Infectious-Recovered (SEIR) models. These models form the foundation for understanding the epidemiological dynamics of a population under the influence of a single-strain pathogen.

In this step, the primary objective is to compute the basic reproduction number, $\gls{r0}$, which quantifies the average number of secondary infections produced by a single infectious individual in a fully susceptible population. $\gls{r0}$ is a critical threshold parameter in disease modeling, as it determines whether a pathogen can invade and establish itself within the host population \cite{dietz1993estimation,macdonald1952analysis,fine1993herd}. If $\gls{r0}>1$, the infection is likely to spread; if $\gls{r0}<1$, the infection will eventually die out \cite{delamater2019complexity,van2008further}.

The mathematical formulation of $\gls{r0}$ depends on the specific structure of the compartmental model. For example, in the SIR model, $\gls{r0}$ is often derived from parameters such as the transmission rate and recovery rate. In models like SEIR, the inclusion of an exposed compartment introduces additional parameters, which must be taken into account in the calculation of $\gls{r0}$ \cite{van2008further}.

This step establishes a foundational understanding of disease dynamics within a single-trait environment (referred to as the resident environment) and without evolutionary processes. It serves as the baseline for assessing the feasibility of mutant strain invasion during the evolutionary process, which is analyzed in subsequent steps.

\subsubsection*{Step 2: Selection of the parameter that undergoes evolution in the base model}
The second step of the framework focuses on identifying the parameter within the epidemiological model that is subject to evolutionary dynamics. This parameter, often called the decision variable, represents an evolvable pathogen trait that changes as a function of forces such as mutation or selection \cite{RN5,boldin2012evolutionary}.

In this study, the decision variable is the \textit{virulence}. Virulence is a critical factor in pathogen evolution, as it can directly influence both the fitness of the pathogen and the dynamics of disease spread through hosts. In some models, virulence can also influence other epidemiological parameters, such as the transmission rate, which can depend on the severity of the disease \cite{van1995dynamics}. Consequently, the choice of the decision variable provides the foundation for linking evolutionary processes to epidemiological outcomes in subsequent analysis steps. This determines the focus of the evolutionary invasion analysis and the traits that will be evaluated for their stability and potential to invade the resident population.

\subsubsection*{Step 3: Categorization of model parameters}

The third step in the evolutionary invasion analysis framework involves systematically categorizing the parameters of the epidemiological model based on their relationship to evolutionary dynamics. This categorization is essential to distinguish between constant parameters and those that evolve or are influenced by evolutionary changes in the decision variable. The parameters can be divided into the following categories:

\begin{itemize}
    \item \textbf{Independent parameters}: These parameters are not influenced by the evolutionary process and remain fixed throughout the analysis. Examples may include birth rates, natural death rates, or environmental factors that do not change with the evolution of the pathogen.
    
    \item \textbf{Evolution-dependent parameters (indirect)}: These parameters are influenced by evolutionary dynamics but are not directly linked to the decision variable. For example, factors such as host immunity levels can indirectly affect the pathogen's evolution but are not themselves functions of the decision variable.
    
    \item \textbf{Parameters as functions of the decision variable}: These parameters are explicitly dependent on the decision variable chosen in Step 2. For example, if virulence is the decision variable, the transmission rate can be modeled as a function of virulence. This relationship may reflect trade-offs or biological constraints, such as the potential for higher virulence to reduce transmission efficiency due to increased host mortality.
\end{itemize}

This step provides a structured approach to understanding how each parameter contributes to the dynamics of the system and the evolutionary processes under study. By clearly identifying the dependencies and relationships among parameters, we can construct more accurate models that capture the interplay between ecological and evolutionary factors.

\subsubsection*{Step 4: Extend the compartmental model to include mutant strains}

This step extends the compartmental epidemiological model to incorporate the infected population carrying a mutant strain. The extended model assumes that a single-strain pathogen has already invaded and established itself within the host population. Additionally, it is assumed that there is no possibility of superinfection, i.e. hosts cannot be simultaneously infected with both resident and mutant strains.

The next task is to identify the \gls{mfe}, denoted as $\hat{V}$, of the extended model. This equilibrium corresponds to the state where the resident strain persists in the population, while the mutant strain remains absent. The stability of this equilibrium is analyzed using the stability criteria for the \gls{mfe}. To facilitate this analysis, a function analogous to the basic reproduction number, $\mathcal{R}_0$, is defined. This function takes the form:
\[
\mathcal{R}_0(\hat{V}) = \frac{\phi(x_m)}{\phi(x_r)},
\]
where $\phi$ represents a function of the decision variable $x$, and $x_m$ and $x_r$ denote the values of the decision variable for the mutant and resident strains, respectively.

The stability criterion for the \gls{mfe}, based on the computed $\mathcal{R}_0(\hat{V})$, serves as a critical threshold parameter in evolutionary invasion analysis. It determines whether a mutant strain can invade and establish itself within the host population:
\begin{itemize}
    \item If $\mathcal{R}_0(\hat{V}) > 1$, the mutant strain is likely to invade and persist within the host population.
    \item If $\mathcal{R}_0(\hat{V}) < 1$, the mutant strain fails to invade.
\end{itemize}

The functional form of $\mathcal{R}_0(\hat{V}) = \frac{\phi(x_m)}{\phi(x_r)}$ communicates that a mutant strain can invade successfully only if $\phi(x_r)$ is not a maximum of the function $\phi$. In other words, the invasion criterion emphasizes the role of evolutionary fitness landscapes in determining whether the mutant strain can outcompete the resident strain. This step establishes the critical threshold for assessing the evolutionary stability of traits in the host-pathogen system.

\subsubsection*{Step 5: Identify the \glspl{ess}}

In this step, the focus shifts to the analysis of the function $\phi(x)$, derived in the previous step, where $x$ represents the decision variable associated with a given strain of the pathogen. The function $\phi(x)$ encapsulates the relationship between the decision variable and the fitness of the strain and serves as the basis for determining the conditions under which evolutionary stability is achieved.

To identify \glspl{ess}, the following process is carried out:
\begin{itemize}
    \item Compute the critical points of the function $\phi(x)$ by solving for the values of $x$ where $\frac{d\phi(x)}{dx} = 0$.
    \item Analyze the second derivative, $\frac{d^2\phi(x)}{dx^2}$, at these critical points to distinguish maxima from minima. The maxima of $\phi(x)$ correspond to the potential \glspl{ess}, as they represent trait values that cannot be invaded by mutant strains with alternative values of $x$.
\end{itemize}

The identified maxima provide the \glspl{ess}, which are trait values for the pathogen that are evolutionarily stable. At these points, no mutant strain with a slightly different trait value can achieve a higher fitness than the resident strain. This step is a pivotal aspect of evolutionary invasion analysis, as it determines the long-term evolutionary outcomes in the host-pathogen system.

\subsubsection*{Step 6: Sensitivity analysis of the \gls{ess}}

The final step in the evolutionary invasion analysis framework involves assessing the sensitivity of the \gls{ess}, denoted as $x^*$, to variations in an external parameter, $y$. This analysis sheds light on how changes in environmental or system parameters influence the stability and selection of traits.

The process begins by considering the function $\phi(x, y)$, where $x$ represents the decision variable and $y$ is the parameter of interest. The sensitivity of the \gls{ess} to $y$ is determined by analyzing the derivatives of $x^*$ with respect to $y$. By
\begin{equation}\label{eq:inva}
    \frac{d x^*}{d y} = - \frac{\frac{\partial^2 \phi(x,y)}{\partial y \partial x}\Big|_{x=x^*}}{\frac{\partial^2 \phi(x,y)}{\partial^2 x}\Big|_{x=x^*}}\propto \frac{\partial^2 \phi(x,y)}{\partial y \partial x}\Big|_{x=x^*},
\end{equation}
the sensitivity analysis can be conducted by examining the sign of the partial derivatives of $\phi(x, y)$ with respect to $x$ and $y$, evaluated at $x = x^*$. 

This step provides critical insight into the robustness and adaptability of the \gls{ess} under varying conditions. It highlights how external factors, such as environmental changes or control interventions, may alter the evolutionary dynamics of the system, potentially leading to shifts in the traits selected as evolutionarily stable. The results of this analysis are vital to designing effective strategies to influence the evolution of pathogens or mitigate the impacts of diseases.

\subsection{Epidemiological models with and without asymptomatic compartments}
We commence our investigation into the evolutionary impact of incorporating an asymptomatic group into epidemiological models by implementing step 1 of the evolutionary invasion analysis framework. In accordance with this step, we focus on epidemiological base models within a single-strain pathogen environment. This analysis involves comparing two models: one that excludes the asymptomatic compartment and another that incorporates an asymptomatic compartment, dividing the infectious population into asymptomatic and symptomatic subgroups.

\begin{figure}[h!]
    \centering
    \includegraphics[width=\textwidth]{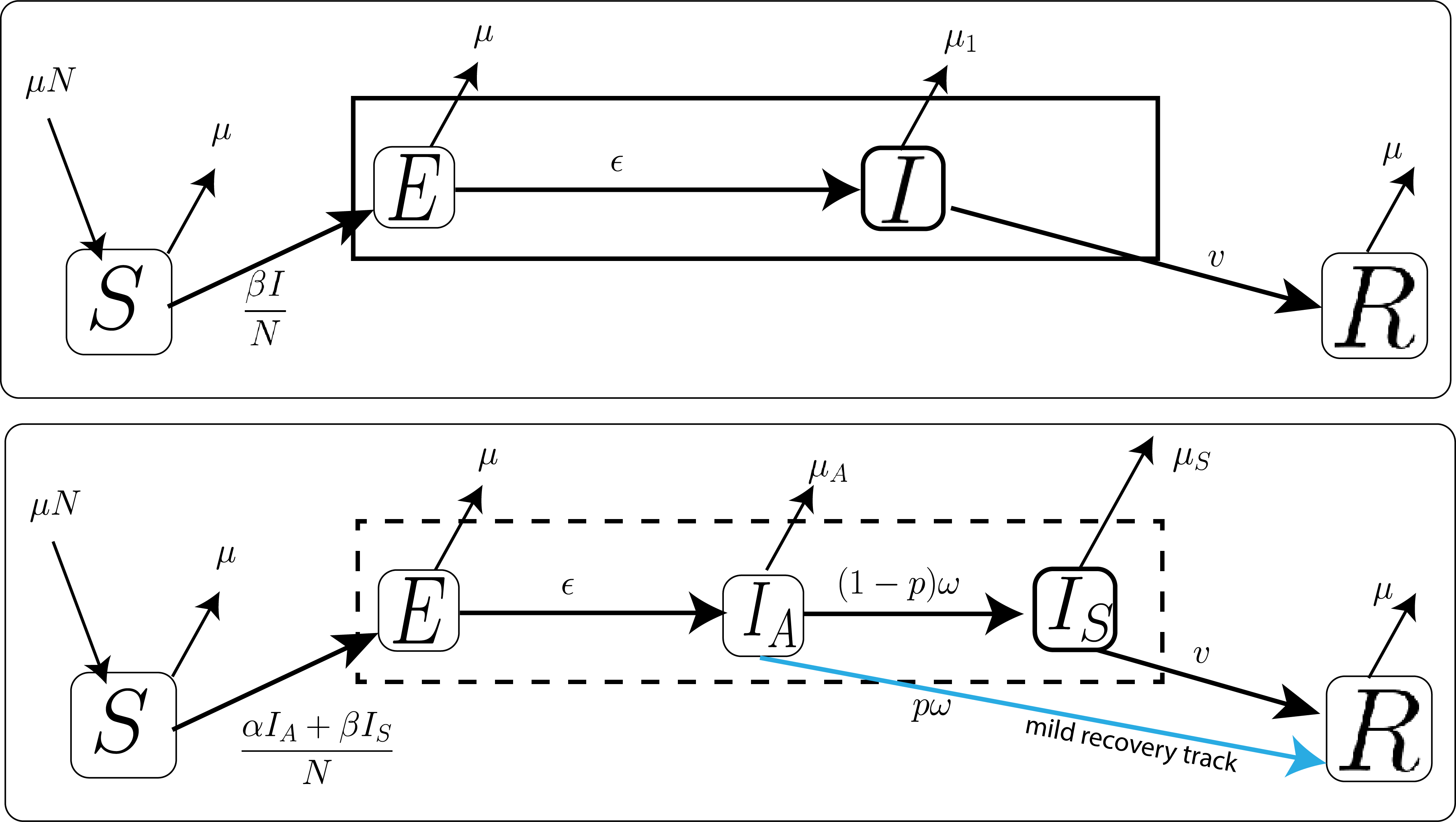}
    \caption{\textbf{System dynamics of SEIR and SEAIR epidemiological models.} An illustration of the system dynamics for the susceptible, exposed, asymptomatic, symptomatic, and recovered compartments in epidemiological models. The top diagram represents the compartmental SEIR model that excludes the asymptomatic population. The bottom diagram depicts the compartmental model incorporating an asymptomatic compartment, where the infectious group is divided into asymptomatic and symptomatic subpopulations. The figure is based on a similar one in a prior manuscript \cite{ogbunugafor2024modes}. Note the mild recovery track (bottom model) that demonstrates how populations of hosts can bypass the symptomatic stage of infection, and move directly toward the recovery compartment.} 
    \label{fig:Models}
\end{figure}

To facilitate this comparison, we describe the SEIR model without an asymptomatic compartment and the extended compartmental model with an asymptomatic compartment (see \cref{fig:Models}). It is important to note that both models represent a single-trait environment (referred to as the resident environment) without evolutionary processes. In \cref{sec:resuls}, we will extend these models to include the effects of a rare mutant pathogen, allowing the evaluation of \glspl{ess} and further exploration of evolutionary invasion dynamics. Our approach aligns with previous studies on the evolution of asymptomatic infection, which consider ``latency" as the decision variable and incorporate two infectious compartments: asymptomatic and symptomatic \cite{saad2020dynamics}. However, unlike previous approaches, we use multivariate virulence parameters associated with asymptomatic and symptomatic compartments as decision variables. This enables the exploration of a two-dimensional decision space for the fitness function.  Furthermore, our model explicitly includes an exposed compartment to distinguish the incubation period from the asymptomatic infectious period. In addition, we incorporate mode-specific parameters to characterize the different transmission routes, including a `mild recovery route, in which asymptomatic individuals can recover without progressing to the symptomatic state.

We now outline the system dynamics of both models using a system of \glspl{ode}. The variables used in these models are as follows: \(N\) represents the total population, while \(S\) denotes the number of susceptible individuals. \(E\) corresponds to the exposed individuals who have been infected but are not yet infectious. \(I\) represents the infectious host population. The infectious population is divided into two subgroups: \(I_A\), which represents asymptomatic individuals, and \(I_S\), representing symptomatic individuals. Finally, \(R\) indicates the number of recovered individuals. All these variables are quantified as the number of people. The parameters utilized in the models are detailed in \cref{tab:Para}.
\begin{table}[h]
\centering
\begin{tabular}{p{2cm}p{7.3cm}p{1.2cm}}
\toprule
\rowcolor{gray!80}
\textbf{Parameters} & \textbf{Description} & \textbf{Units} \\
\midrule
\rowcolor{white}
$\mu$ & Natural death rate (Also use to represent the birth rate)  & $\si{day}^{-1}$  \\
\rowcolor{gray!30}
$\boldsymbol{\mu_1}$ & \textbf{Virulence } (for SEIR-model)&$\boldsymbol{\si{\textbf{day}}^{-1}}$  \\
\rowcolor{white}
$\boldsymbol{\mu_A}$ & \textbf{Asymptomatic infected virulence} &$\boldsymbol{\si{\textbf{day}}^{-1}}$  \\
\rowcolor{gray!30}
$\boldsymbol{\mu_S}$ & \textbf{Symptomatic infected virulence} &$\boldsymbol{\si{\textbf{day}}^{-1}}$  \\
\rowcolor{white}
$\omega^{-1}$ & Expected time in the asymptomatic state  & $\si{days}$  \\
\rowcolor{gray!30}
$v$ & Recovery rate  & $\si{day}^{-1}$  \\
\rowcolor{white}
$p$ & The fraction that moves along the “mild” recovery track &  \\
\rowcolor{gray!30}
$\epsilon^{-1}$ & Average number of days before infectious & $\si{days}$  \\
\rowcolor{white}
$\alpha$ & Transmission rate through the asymptomatic individuals  & $\si{day}^{-1}$  \\
\rowcolor{gray!30}
$\beta$ & Transmission rate through the symptomatic individuals  & $\si{day}^{-1}$  \\

\bottomrule
\end{tabular}
\caption{The parameters used in epidemiological models. In the models, infected individuals who are bedridden and unable to participate in population circulation are considered a measure of virulence. This measure is highlighted in the table with a bold row.}
\label{tab:Para}
\end{table}
\subsubsection{Model 1: without asymptomatic transmission}
This epidemiological model is structured as a classical SEIR compartmental model. The system dynamics of the model are represented by:
\begin{equation}\label{eq:SEIRModel}
    \begin{aligned}
        \frac{d S}{dt} &= \mu (N-S) -\frac{ \beta I}{N}S\\
        \frac{d E}{dt} &=  \frac{ \beta I}{N}S-(\epsilon+ \mu)E \\
        \frac{d I}{dt} &= \epsilon E -(\nu + \mu_1) I \\
        \frac{d R}{dt} &= \nu I - \mu R.
    \end{aligned}
\end{equation}
As part of step 1 of the evolutionary invasion analysis framework, the basic reproduction number, $\mathcal{R}_{0_1}$, for this model can be calculated, and it is given by:
\begin{equation}
    \begin{aligned}
        \mathcal{R}_{0_1}= \frac{\epsilon \beta S^*}{N(\epsilon+\mu)(\nu+\mu_1)}.
    \end{aligned}
\end{equation}
The stability analysis of the system at the \gls{dfe}, defined as \((S^*, E^*, I^*, R^*) = (N, 0, 0, 0)\), can be utilized to derive the basic reproduction number. The computation is straightforward, and the detailed steps can be found in \cite{van2008further}. In this notation, \(\mathcal{R}_{0_1}\), the subscript \(1\) indicate the model number (which indicate the model without the asymptomatic compartment).
\subsubsection{Model 2: with asymptomatic transmission}
To incorporate the asymptomatic infectious population into a epidemiological model, as illustrated in \cref{fig:Models}, the infectious population is divided into two subpopulations, represented by the asymptomatic and symptomatic compartments. The complete model is described by the system of \glspl{ode} given by:  
\begin{equation}\label{eq:covidModel}
    \begin{aligned}
        \frac{d S}{dt} &= \mu (N-S) -\big( \frac{\alpha I_A + \beta I_S}{N}\big)S\\
        \frac{d E}{dt} &= \big( \frac{\alpha I_A + \beta I_S}{N}\big)S-(\epsilon + \mu)E \\
        \frac{d I_A}{dt} &= \epsilon E -(\omega + \mu_A) I_A\\
        \frac{d I_S}{dt} &= (1-p)\omega I_A -(v + \mu_S) I_S\\
        \frac{d R}{dt} &= p\omega I_A+v I_S - \mu R.
    \end{aligned}
\end{equation}
It is important to note that the \gls{dfe} of the system simplifies to \((S^*, E^*, I_A^*, I_S^*, R^*) = (N, 0, 0, 0, 0)\). The next generation matrix (NGM) approach \cite{diekmann1990definition, diekmann2010construction} provides a method to derive the basic reproduction number for this model. This approach entails linearizing the infected subsystems of the \glspl{ode} around the \gls{dfe} and representing them in matrix form as:  
\begin{equation}  
    \frac{d\mathbf{X}}{dt} = \mathbf{(F - V)X},  
\end{equation}  
where $\mathbf{X} =\begin{pmatrix}
    E \\
    I_A \\
    I_S
\end{pmatrix}$. The matrix \(\mathbf{F}\) encapsulates the transmission dynamics, representing all flows from uninfected to infected states, while the matrix \(\mathbf{V}\) accounts for all other transitions within the system \cite{castillo2020tour, diekmann2010construction, surasinghe2024structural}. Using this approach, the basic reproduction number \(\mathcal{R}^2_{0r}\) for this model can be computed as \(\mathcal{R}_{0_2} = \rho(\mathbf{FV}^{-1})\), where \(\rho(\cdot)\) denotes the spectral radius of a matrix. The matrices \(\mathbf{F}\) and \(\mathbf{V}\) for this model at the disease-free equilibrium are given by:
\begin{equation*}
    \begin{aligned}
        \mathbf{F}=\begin{pmatrix}
            0 &\alpha &\beta\\
            0& 0 & 0 \\
            0 &0& 0
        \end{pmatrix}, \  
        \mathbf{V}=\begin{pmatrix}
            \epsilon+\mu &0 &0\\
            -\epsilon  & \omega + \mu_A & 0 \\
            0 &-(1-p)\omega  & v+\mu_S
        \end{pmatrix} . 
    \end{aligned}
\end{equation*}
Hence, the basic reproduction number, \(\mathcal{R}_{0_2}\) (where the subscript $2$ denotes the model number), for the model can be simplified to:
\begin{equation}\label{eq:covidR0}
\mathcal{R}_{0_2}=\rho(\mathbf{F}\mathbf{V}^{-1})=\frac{\epsilon\big(\alpha(v+\mu_S)+\beta\omega(1-p)\big)}{(\epsilon+\mu)(\omega+\mu_A)(v+\mu_S)}.
\end{equation}

\section{Results}\label{sec:resuls}
In this section, we conduct a detailed evolutionary invasion analysis of models with and without the asymptomatic infectious group to understand the impact of the asymptomatic transmission on the evolutionary dynamics. In this analysis, we follow the step-by-step process outlined in the framework (discussed in \cref{sec:framework}\cite{surasinghe2024evolutionary}). First, we examine Model 1, which represents the system without asymptomatic transmission. Next, we analyze Model 2, incorporating asymptomatic transmission, and compare the results with those from Model 1. For both models, we consider virulence as the decision variable and assume transmission rates to be a function of virulence.
\subsection{Evolutionary invasion analysis for the model without asymptomatic transmission} \label{sec:EinModel1}
In this model, the infected virulence, \(\mu_1\), is considered the decision variable subject to evolutionary dynamics. Additionally, we assume the transmission rate, \(\beta\), to be a function of virulence, with no other parameter influenced by evolution. Accordingly, \(\mu_{1}^j\) denotes the value of virulence for \(j = r, m\), where \(r\) corresponds to the resident strain and \(m\) to the rare mutant strain. For simplicity, we use the shorthand notation \(\beta_j\) to represent the function value, \(\beta_j = \beta(\mu_1^j)\). The next step in the process involves extending the epidemiological model to incorporate the presence of a rare mutant pathogen in the environment. This step requires evaluating the \gls{mfe}, \(\hat{V}_1\), and determining its stability condition to compute \(\gls{r0}_1(\hat{V}_1)\). The extended model, where \(E_j\) and \(I_j\) represent the exposed and infectious populations for the strain \(j = r, m\), is given by: 
\begin{equation}\label{eq:SEIRModelEv}
    \begin{aligned}
        \frac{d S}{dt} &= \mu (N-S) -\frac{ \beta_r I_r}{N}S-\frac{ \beta_m I_m}{N}S\\
        \frac{d E_r}{dt} &=  \frac{ \beta_r I_r}{N}S-(\epsilon+ \mu)E_r \\
        \frac{d E_m}{dt} &=  \frac{ \beta_m I_m}{N}S-(\epsilon+ \mu)E_m \\
        \frac{d I_r}{dt} &= \epsilon E_r -(\nu + \mu_1^r) I_r \\
        \frac{d I_m}{dt} &= \epsilon E_m -(\nu + \mu_1^m) I_m \\
        \frac{d R}{dt} &= \nu I_r+\nu I_m - \mu R.
    \end{aligned}
\end{equation}
Following the computational steps outlined in \cite{surasinghe2024evolutionary}, \(\gls{r0}_1(\hat{V}_1)\) for this model is derived as:
\begin{equation}
    \gls{r0}_1(\hat{V})= \frac{\epsilon \beta_m \hat{S}}{N(\epsilon+\mu)(\nu+\mu_1^m)}.
\end{equation}
where \(\hat{S}\) represents the susceptible (\(S\)) population at the \gls{mfe}, and it is given by:
\begin{equation}
    \hat{S}= \frac{N(\epsilon+\mu)(\nu+\mu_1^r)}{\epsilon \beta_r }.
\end{equation}
Hence, \(\gls{r0}_1(\hat{V}_1)\) can be expressed as \(\gls{r0}_1(\hat{V}_1) = \frac{\mathcal{R}_{0_1}^m}{\mathcal{R}_{0_1}^r}\), where \(\mathcal{R}_{0_1}^j\) denotes the basic reproduction number for model 1(\cref{eq:SEIRModel}) with strain \(j (=r,m)\) and can be interpreted as the fitness function for the strain. However, in this article, since we assume that the parameters \(\epsilon\) and \(\mu\) remain unchanged by the evolution, the value of \(\gls{r0}_1(\hat{V})\) can be further simplified to \(\gls{r0}_1(\hat{V}_1) = \frac{\phi_1(\mu_1^m)}{\phi_1(\mu_1^r)}\), where the fitness function $\phi_1(\cdot)$ is given by:
\begin{equation} \label{eq:fitModel1}
    \phi_1(\mu_1)=\frac{\beta(\mu_1)}{\nu + \mu_1}
\end{equation}

\subsubsection{\gls{ess} and sensitivity analysis of \gls{ess} for the model without asymptomatic transmission}
The ``invasive fitness" \(\phi_1\), as defined in \cref{eq:fitModel1}, represents the fitness of a rare mutant allele emerging within a population dominated by a resident allele for the model without asymptomatic transmission. If the resident strain is at its \gls{ess}, the fitness function attains its maximum value at that point. Consequently, the \gls{ess}, \(\mu_1^*\), for the model is determined by:
\begin{equation}
    \mu_1^* = \argmax_{\mu_1} \phi_1(\mu_1).
\end{equation}
The first and second derivative tests for maxima can be utilized to explore the conditions that must be satisfied by the \gls{ess}. By setting the first derivative of the fitness function to zero and ensuring that the second derivative is negative, the \gls{ess}, \(\mu_1^*\), must satisfy the following conditions:
\begin{equation}\label{eq:trasCriForModel1}
    \beta'(\mu_1^*)=\frac{\beta(\mu_1^*)}{\nu+\mu_1^*},\ \text{ and }\  \beta''(\mu_1^*)<0
\end{equation}
where the prime ($'$) and double prime ($'$) denote the first and second derivatives of the function with respect to $\mu_1$. The final conditions reduce to constraints on the transmission rate $\beta$ at the \gls{ess}, $\mu_1^*$. Since all parameters are assumed to be positive, these conditions can be interpreted as follows: The first derivative of the transmission function at $\mu_1^*$ must be positive, and the second derivative must be negative at $\mu_1^*$. Thus, an \gls{ess} is possible only if the transmission rate increases at a diminishing rate as virulence around \(\mu_1^*\). In other words, the transmission rate must be increasing and concave down at a \gls{ess}. For demonstration purposes, we will use a transmission function of the form \(\beta(\mu_1) = a \sech^2(b\mu_1 - c)\) and select the hyperparameters \(a\), \(b\), and \(c\) such that they satisfy the conditions discussed above. 

Using the criteria provided in \cref{eq:inva}, the sensitivity of the recovery rate to the \gls{ess} can be evaluated. For this model, the corresponding criterion is given by: 
\begin{equation}
    \frac{d\mu_1^*}{d\nu} \propto \frac{\beta'(\mu_1^*)}{(\nu+\mu_1^*)^2}>0.
\end{equation}
Hence, the \gls{ess} level of virulence ( $\mu_1^*$ ) will always increase as the recovery rate ($\nu$) increases, regardless of the exact relationship between the transmission rate and virulence, provided it satisfies \cref{eq:trasCriForModel1}. Note that for interpretation, \( \nu^{-1} \) can be considered, as it represents the mean infectious period (the expected time spent in the infectious (\(I\)) state). Thus, the previous conclusion can be interpreted as follows: the \gls{ess} level of virulence will decrease as the mean infectious period increases.

\subsection{Evolutionary invasion analysis for a model with asymptomatic transmission}
Similar to the analysis in \cref{sec:EinModel1}, we extend the single-strain model with asymptomatic transmission, outlined in \cref{eq:covidModel}, to incorporate the presence of a rare mutant pathogen in the environment. As mentioned previously, we assume no superinfection. Consequently, the subpopulation influenced by the pathogen is further divided into \(E_j\), \(I_{A_j}\), and \(I_{S_j}\), representing the exposed, asymptomatic infectious, and symptomatic infectious populations with pathogen strain \(j = r\) (resident) or \(j=m\) (mutant), respectively. We assume that both virulence, \(\mu_A\) and \(\mu_S\), in the asymptomatic and symptomatic populations due to the pathogen are subject to evolutionary changes, denoted as \(\mu_A^j\) and \(\mu_S^j\) for \(j = r, m\), where \(r\) represents the resident strain and \(m\) represents the mutant strain. Furthermore, the transmission rates in both groups are also considered to vary as a result of the pathogen's evolution, denoted as \(\alpha_j\) and \(\beta_j\) for \(j = r, m\). The extended two-strain model is then formulated as follows: 

\begin{equation}\label{eq:covidModelEv}
    \begin{aligned}
        \frac{d S}{dt} &= \mu (N-S) -\big( \frac{\alpha_r I_{A_r} + \beta_r I_{S_r}}{N}\big)S - \big( \frac{\alpha_m I_{A_m} + \beta_m I_{S_m}}{N}\big)S\\
        \frac{d E_j}{dt} &= \big( \frac{\alpha_j I_{A_j} + \beta_j I_{S_j}}{N}\big)S-(\epsilon + \mu)E_j &\text{ for } j=r,m \\
        \frac{d I_{A_j}}{dt} &= \epsilon E -(\omega + \mu_A^j) I_{A_j} &\text{ for } j=r,m\\
        \frac{d I_{S_j}}{dt} &= (1-p)\omega I_{A_j} -(\nu + \mu_S^j) I_{S_j} &\text{ for } j=r,m\\
        \frac{d R}{dt} &= p\omega (I_{A_r}+I_{A_m})+\nu (I_{S_r}+I_{S_m}) - \mu R.
    \end{aligned}
\end{equation}

This model incorporates the dynamics of both resident and mutant strains, accounting for evolutionary changes in virulence \(\mu_A\) and \(\mu_S\), as well as transmission rates for asymptomatic and symptomatic populations. By including these aspects, the model provides a comprehensive framework for analyzing the impact of evolutionary adaptations on asymptomatic transmission. To derive the invasion fitness function for this model, we will employ the computational steps outlined in \cite{surasinghe2024evolutionary}. 

The susceptible population, \(\hat{S}\), at the \gls{mfe} ($\hat{V}_2$) for this model is expressed as: 
\begin{equation}\label{eq:CovidSr}
\hat{S}= \frac{N(\epsilon+\mu)(\omega+\mu_A^r)(v+\mu_S^r)}{\epsilon\big(\alpha_r(v+\mu_S^r)+\beta_r\omega(1-p)\big)}.
\end{equation}
Furthermore, it is noted that the basic reproduction number for the \gls{mfe} (\(\gls{r0}_2(\hat{V}_2)\)) can be expressed as the following equation: 
\begin{equation} \label{eq:covidR0DFE}
\gls{r0}_2(\hat{V}_2)=\frac{\hat{S}\epsilon\big(\alpha_m(v+\mu_S^m)+\beta_m\omega(1-p)\big)}{N(\epsilon+\mu)(\omega+\mu_A^m)(v+\mu_S^m)}.
\end{equation}
By rearranging the terms of the equation, \(\gls{r0}_2(\hat{V}_2)\) can be simplified to 
\(\gls{r0}_2(\hat{V}_2) = \frac{\phi_2(\mu_A^m, \mu_S^m)}{\phi_2(\mu_A^r, \mu_S^r)}\), 
where the invasion fitness function \(\phi_2 : [0,1] \times [0,1] \to \mathbb{R}\) is given by: 
\begin{equation}
\phi_2(\mu_A, \mu_S) = \frac{\alpha}{\omega+\mu_A}+ \frac{\omega(1-p)}{\omega+\mu_A} \Bigg( \frac{\beta}{v+\mu_S}\Bigg).
\end{equation}
We further assume that the transmission rates \(\alpha = \alpha(\mu_A)\) and \(\beta = \beta(\mu_S)\) are functions of \(\mu_A\) and \(\mu_S\), respectively. For simplicity, we use the shorthand \(\alpha, \beta\) to represent these functions throughout the discussion. Additional assumptions, such as \(\mu_A = \mu\) remaining unchanged with evolution or the transmission rate \(\alpha\) being unaffected by evolutionary changes, can be used to simplify the analysis. Detailed exploration of these simplified cases can be found in \cite{surasinghe2024evolutionary}. However, in this article, we focus on exploring the \gls{ess} of two variables, \(\mu_A\) and \(\mu_S\). Consequently, the fitness function considered here is a multivariable function.

\subsubsection{\gls{ess} and sensitivity analysis of \gls{ess} for the model with asymptomatic transmission}
To identify the \glspl{ess}, the maxima of the fitness function and the corresponding values that achieve these maxima must be determined. Since the fitness function in this model is a multivariable function, vector calculus techniques are required. Specifically, the first derivative (gradient) is utilized to identify the critical points, while the second derivative (Hessian matrix) test is employed to classify these critical points as maxima. These methods are instrumental in determining the \glspl{ess} for this model.

By setting the first derivatives, \(\frac{\partial \phi_2}{\partial \mu_A}\) and \(\frac{\partial \phi_2}{\partial \mu_S}\), of the fitness function to zero and ensuring that the second derivative test confirms the maxima at the critical points, the \gls{ess}, \((\mu_A^*, \mu_S^*)\), must satisfy the following conditions:  
\begin{equation}\label{eq:ConModel2}
\begin{aligned}
    \beta'(\mu_S^*)&=\frac{\beta(\mu_S^*)}{v+\mu_S^*}, \\
    \alpha'(\mu_A^*) &= \phi_2(\mu_A^*,\mu_S^*), \\
    \text{and} \quad &\beta''(\mu_S^*), \ \alpha''(\mu_A^*) > 0,
\end{aligned}
\end{equation}
where the prime ($'$) and double prime ($''$) denote the first and second derivatives of a single-variable function with respect to that variable. Similar to the SEIR model, the final conditions reduce to constraints on the transmission rates \(\alpha\) and \(\beta\) in the\gls{ess}, \((\mu_A^*, \mu_S^*)\). Since all parameters are assumed to be positive, these conditions can be interpreted as follows: the first derivatives of the respective transmission functions (\(\alpha, \beta\)) at \(\mu_A^*\) and \(\mu_S^*\) must be positive, while the second derivatives must be negative at those points. Thus, an \gls{ess} is feasible only if the transmission rates (\(\alpha, \beta\)) increase at a diminishing rate as virulence approaches \((\mu_A^*, \mu_S^*)\). In other words, transmission rates must exhibit an increasing behavior while being concave down at the \gls{ess}. 

For the purpose of conducting a sensitivity analysis with a given parameter \(y\), and analogous to \cref{eq:inva}, it can be demonstrated that, under a multi-evolutionary variable scenario, the following relationships hold:  
\begin{equation}
\frac{d \mu_A^*}{dy} \propto \frac{\partial^2 \phi_{2}}{\partial y \partial\mu_A}\Big|_{(\mu_A^*,\mu_S^*)} \text{ and } \frac{d \mu_S^*}{dy} \propto \frac{\partial^2 \phi_{2}}{\partial y \partial\mu_S}\Big|_{(\mu_A^*,\mu_S^*)}.
\end{equation}
Hence, the sensitivity analysis of parameter \(y\) can be conducted by examining the sign of the partial derivatives of \(\phi_2(\mu_A, \mu_S, y)\) with respect to virulence (\(\mu_A\) or \(\mu_S\)) and \(y\), evaluated at the \gls{ess} \((\mu_A^*, \mu_S^*)\). For this model, we can choose the sensitivity analysis parameter \(y\) as \(\nu\), \(\omega\), and \(p\).

Note that,  
\begin{equation} 
\frac{d \mu_A^*}{d\nu}\propto \frac{(1-p)\omega\beta(\mu_S^*)}{(\omega+\mu_A^*)^2(\nu+\mu_S^*)^2}>0 \text{ and } \frac{d \mu_S^*}{d\nu}\propto \frac{(1-p)\omega\beta'(\mu_S^*)}{(\omega+\mu_A^*)(\nu+\mu_S^*)^2}>0. 
\end{equation}  
Therefore, both virulence values at the \gls{ess}, \(\mu_A^*\) and \(\mu_S^*\), increase as the recovery rate increases.
When considering the partial derivatives with respect to \(\omega\), the following relationships hold:  
\begin{equation}
    \frac{d \mu_A^*}{d\omega}\propto \frac{1}{(\omega+\mu_A^*)^3} \Bigg(\alpha(\mu_A^*)-(1-p)\mu_A^*\frac{\beta(\mu_S^*)}{\nu+\mu_S^*} \Bigg) \text{ and } \frac{d \mu_S^*}{d\omega}=0.
\end{equation}  
Therefore, \(\mu_A^*\) increases as \(\omega\) increases only if the condition \( \frac{\beta(\mu_S^*)}{\nu+\mu_S^*} < \frac{\alpha(\mu_A^*)}{(1 - p) \mu_A^*}\) is satisfied. However, \(\mu_S^*\) remains unchanged with variations in \(\omega\).  Considering the partial derivatives with respect to \(p\), the following relationship holds:

\begin{equation}
    \frac{d \mu_A^*}{d p}\propto \frac{\omega \beta(\mu_S^*)}{(\omega+\mu_A^*)^2(\nu+\mu_S^*)}>0 \text{ and } \frac{d \mu_S^*}{dp}=0.
\end{equation}
Hence, \(\mu_A^*\) increases as \(p\) increases, while \(\mu_S^*\) remains unchanged with respect to changes in \(p\).

\subsection{Comparison of the two models}\label{sec:ComModel}
The evolutionary dynamics of the pathogen in model 2 (with asymptomatic transmission) are characterized by two major variables. The first involves the virulence of the symptomatic group (\(\mu_S\)) and the transmission rate in that compartment (\(\beta(\mu_S)\)). This variable mirrors the behavior observed in model 1 (SEIR: without asymptomatic transmission). The second evolutionary variable, which introduces interesting evolutionary dynamics, relates to the virulence (\(\mu_A\)) and the transmission rate (\(\alpha(\mu_A)\)) of the asymptomatic group. The pathogen can optimize its fitness by modifying this virulence, thereby influencing asymptomatic transmission. Consequently, this evolutionary variable is sensitive to two additional key factors: the ``expected time in the asymptomatic state" (\(\omega^{-1}\)) and the ``fraction that follows the mild recovery track" (\(p\)).  


Hence, we can summarize (see \cref{tab:Sumaary}) the \gls{ess} and sensitivity analysis as a comparison between symptomatic and asymptomatic transmission. Symptomatic transmission refers to the evolutionary behavior of the pathogen in the SEIR model or the symptomatic group in model 2 (characterized by virulence \(\mu_S\) or \(\mu_1\) and transmission rate \(\beta\)). In contrast, asymptomatic transmission captures the evolutionary dynamics of the pathogen within the asymptomatic group in model 2.  
\begin{table}[h]
\centering
\begin{tabular}{p{3.5cm}|p{3.75cm}|p{3cm}|p{5.5cm}}
\toprule
\rowcolor{gray!80}
\textbf{Description} & \textbf{Real world context}&\textbf{Model 1 (No asymptomatic   transmission)}  & \textbf{Model 2 (with asymptomatic   transmission)} \\
\midrule
\rowcolor{white}
Fitness function & &$\phi_1(\mu_1)=\frac{\beta(\mu_1)}{\nu + \mu_1}$.  & $\phi_2(\mu_A, \mu_S) = \frac{\alpha}{\omega+\mu_A}+ \frac{\omega(1-p)\phi_1(\mu_S)}{\omega+\mu_A}$.  \\
\rowcolor{gray!30}
\gls{ess} given by  & &$\beta'(\mu_1^*)=\frac{\beta(\mu_1^*)}{\nu+\mu_1^*}$.&$\alpha'(\mu_A^*) = \phi_2(\mu_A^*,\mu_S^*)$ and $\beta'(\mu_S^*)=\frac{\beta(\mu_S^*)}{\nu+\mu_S^*}$. \\
\rowcolor{white}
 \multicolumn{4}{c}{\textbf{Sensitivity Analysis}} \\
\rowcolor{gray!30}
Increase of expected time spent in the
symptomatic state ($\nu^{-1}$) & When susceptible individuals are especially vulnerable, have low access to healthcare, or are in poor health &The \gls{ess} value of $\mu_1^*$ decreases. & Both \gls{ess} values of $\mu_A^*$ and $\mu_S^*$ decrease.  \\
\rowcolor{white}
Increase of expected time in the asymptomatic state ($\omega^{-1}$)& When susceptible individuals shed infectious pathogens for longer periods of time &Not applicable & The \gls{ess} value of $\mu_A^*$ decreases only if  $\phi_1(\mu_S^*) < \frac{\alpha(\mu_A^*)}{(1 - p) \mu_A^*}$. The \gls{ess} value of $\mu_S^*$ remains unchanged.  \\
\rowcolor{gray!30}
Increase of the fraction that moves along the ``mild” recovery track ($p$) & When individuals rapidly recover, or when public health interventions steer asymptomatics toward full pathogen clearance &Not applicable  & The \gls{ess} value of $\mu_A^*$ increases. The \gls{ess} value of $\mu_S^*$ remains unchanged.  \\

\bottomrule
\end{tabular}
\caption{The comparison of the \gls{ess} and sensitivity analysis for symptomatic (model 1 or the symptomatic group in model 2)  and asymptomatic (asymptomatic group in model 2) transmission. Note the inclusion of example contexts where the model particulars manifest in real world settings }
\label{tab:Sumaary}
\end{table}

Analyzing the fitness function $\phi_2(\mu_A, \mu_S)$ with asymptomatic transmission, we observe that it consists of two terms. The first term, $\frac{\alpha}{\omega + \mu_A}$, resembles the fitness function for symptomatic transmission, $\phi_1(\mu_S) = \frac{\beta}{\nu + \mu_S}$, but with parameters corresponding to asymptomatic transmission. If each host exclusively follows the `mild' (asymptomatic) recovery path ($p = 1$), model 2 reduces to model 1 (SEIR) with virulence $\mu_A$, transmission rate $\alpha$, and recovery rate $\omega$. Furthermore, in this case, the second term vanishes, and $\phi_2$ simplifies to the first term. Hence, considering $\mu_A$ as an evolutionary variable is crucial, and its evolutionary behavior in this scenario corresponds to the virulence of the infectious population. Therefore, when $p$ increases, the symptomatic infectious population decreases, while the asymptomatic infectious population increases. Consequently, the evolutionary behavior of the pathogen in the asymptomatic population tends to align with that of the symptomatic or common infectious group.
 
 The second term, $\frac{\omega(1-p)}{\omega+\mu_A} \Bigg( \frac{\beta}{v+\mu_S}\Bigg)$, of the fitness function $\phi_2$ accounts for the evolutionary behavior of the pathogen in the symptomatic host environment. It consists of two parts: the fitness $\phi_1(\mu_S)$ of the pathogen in the symptomatic host environment and the rate of production of symptomatic hosts, $\frac{(1 - p) \omega}{\omega + \mu_A} < 1$. Hence, the contribution of the pathogen's fitness in the symptomatic host to the overall fitness $\phi_2$ of model 2 is a fraction of the fitness $\phi_1(\mu_S)$ in model 1. Also, note that as $p$ increases, this contribution decreases, and the highest contributing factor is $\frac{\omega}{\omega + \mu_A}$ when $p = 0$ (i.e., when every host becomes symptomatic after staying in the asymptomatic state).

\subsubsection{Numerical example}\label{esc:numExam}
In this section, we demonstrate the impact of asymptomatic transmission by comparing the models using a numerical example. This analysis considers the evolutionary variable $\mu_1$ for Model 1 and $\mu_A, \mu_S$ for Model 2. The parameter values \cite{ogbunugafor2020variation} are chosen as $\mu=0.000034$, $\omega^{-1}=3.119$, $v=0.031$, $p=0.956$ and $\epsilon^{-1}=2.381$. The transmission rate $\beta$ in model 1 is assumed to be a function of the virulence $\mu_1$ and is defined as $\beta(\mu_1) = a \sech^2(b\mu_1 - c)$. The hyperparameters \(a\), \(b\), and \(c\) must be selected to satisfy the conditions outlined in \cref{eq:trasCriForModel1}. In model 2, the transmission rate $\beta$ for the symptomatic group is assumed to be a function of the virulence $\mu_S$, given by $\beta(\mu_S) = a_1 \sech^2(b_1\mu_S - c_1)$. Similarly, the transmission rate $\alpha$ for the asymptomatic group in model 2 is assumed to be a function of the virulence $\mu_A$, expressed as $\alpha(\mu_A) = a_2 \sech^2(b_2\mu_A - c_2)$. Additionally, it is necessary to select the hyperparameters \(a_i\), \(b_i\), and \(c_i\) for \(i = 1, 2\) such that the conditions outlined in \cref{eq:ConModel2} are satisfied. 

\begin{figure}[htbp]
    \centering
    \begin{subfigure}[b]{0.46\textwidth}
        \centering
        \includegraphics[width=\textwidth]{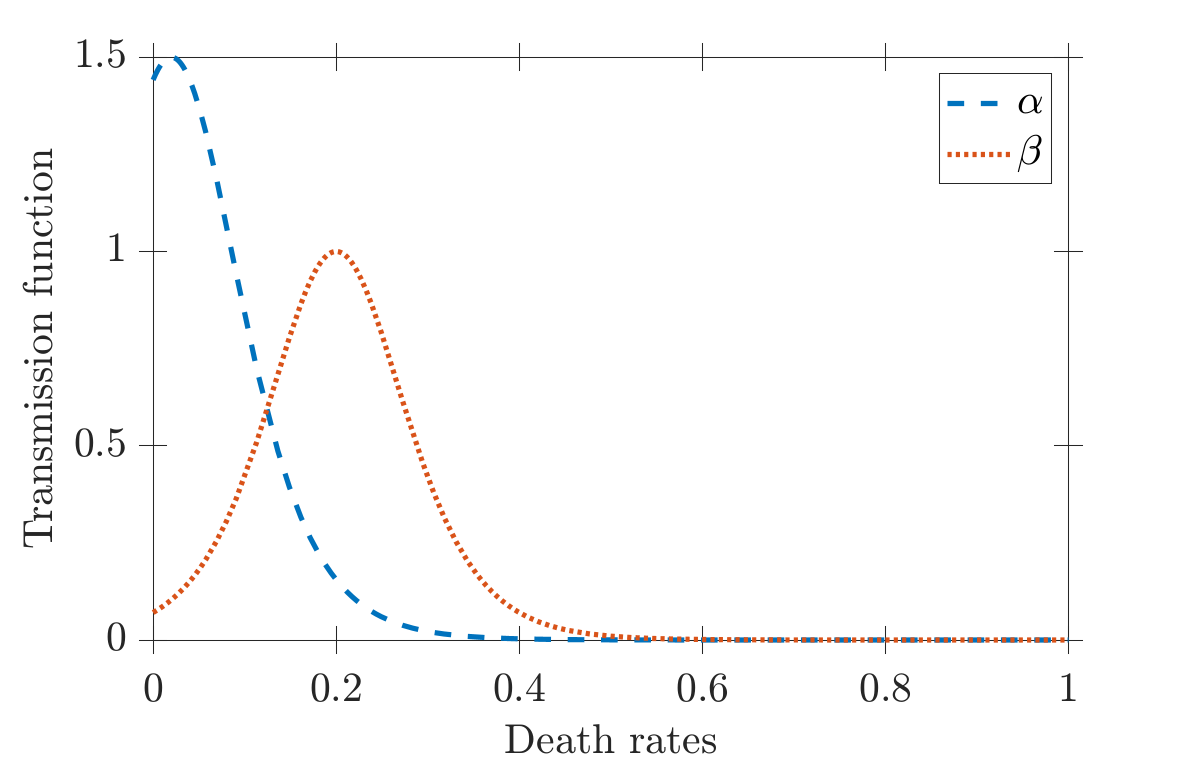}
        \caption{Transmission functions}\label{fig:transmis}
    \end{subfigure}
    \begin{subfigure}[b]{0.47\textwidth}
        \centering
        \includegraphics[width=\textwidth]{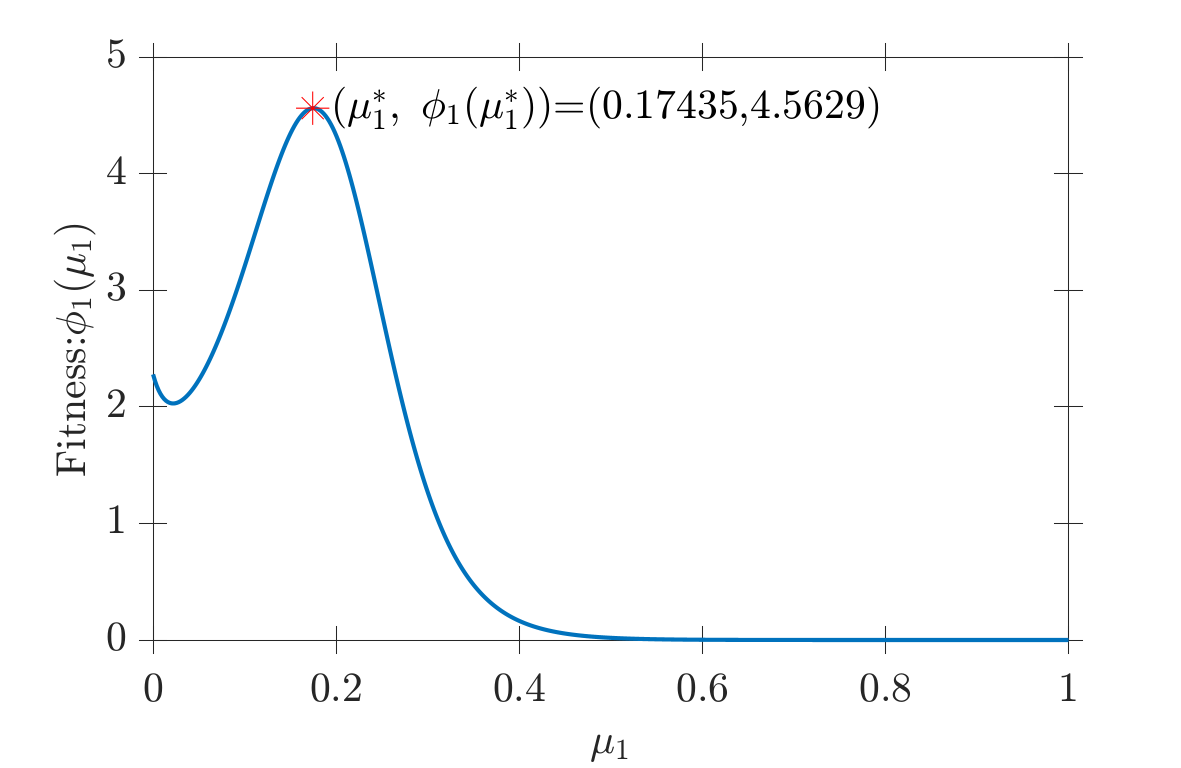}
        \caption{Fitness function of model 1.}\label{fig:ftness1}
    \end{subfigure}
    \caption{\textbf{Transmission functions and fitness function.} (a) Transmission functions for both models and (b) the fitness function $\phi_1$ for Model 1. The chosen transmission functions are $\alpha(x) = 1.5\sech^2(10x - 0.2)$ and $\beta(x) = \sech^2(10x - 2)$, corresponding to the respective virulence $x$. The fitness function $\phi_1$ for Model 1, with the selected transmission function $\beta$ and parameter values, attains its maximum at $\mu_1^* = 0.17435$, which represents the \gls{ess} value for Model 1 under the given parameter configuration.}
\end{figure}

We select the hyperparameters as $a = a_2 =1$, $b = b_2 =10$, $c = c_2 =2$ and $a_1 =1.5$, $b_1 =10$, $c_1 =0.2$ to ensure that model 1 and the symptomatic group share the same transmission function for virulence $\mu_1$ or $\mu_S$. Additionally, the asymptomatic transmission rate is higher than that of the symptomatic group, with the asymptomatic group exhibiting a lower virulence than the symptomatic group (see \cref{fig:transmis}).
Note that by setting the SEIR model's transmission rate equal to the symptomatic transmission rate, we assume that the SEIR model considers only symptomatic individuals as the infectious group.

\begin{figure}[htbp]
    \centering
    \begin{subfigure}[b]{0.8\textwidth}
        \centering
        \includegraphics[width=\textwidth]{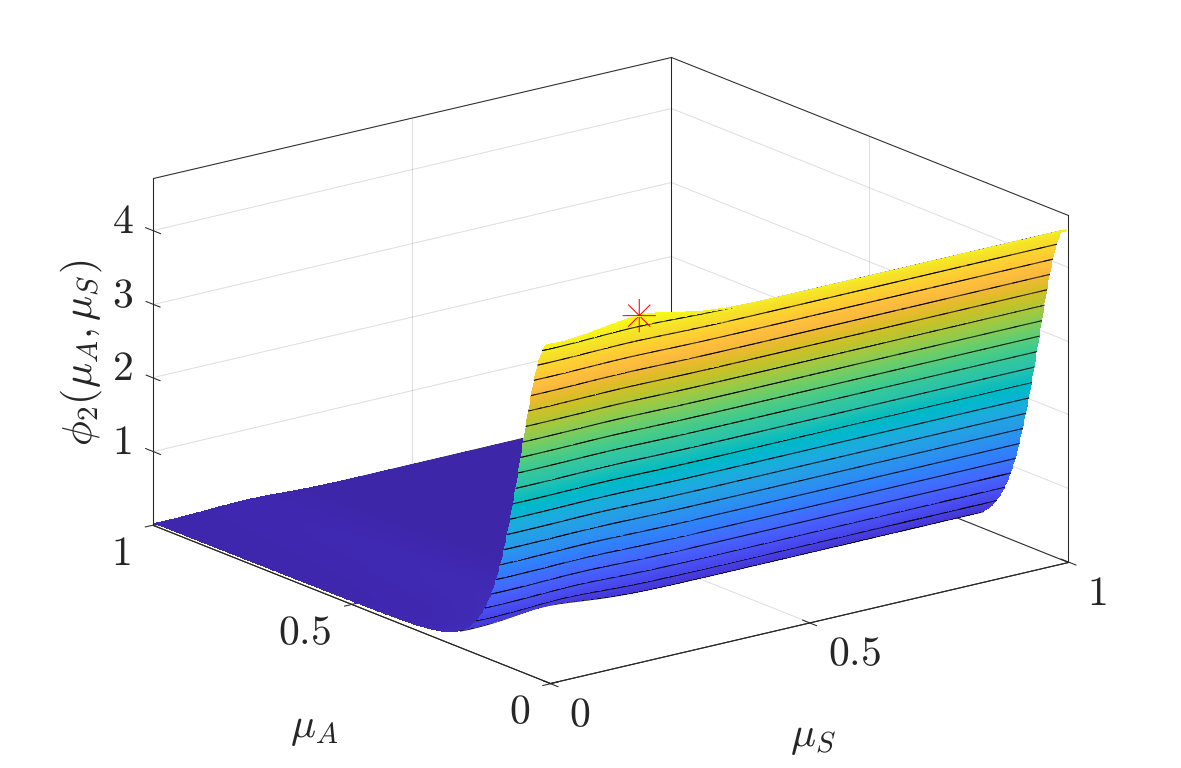}
        \caption{}\label{fig:2dfigphi2}
    \end{subfigure}
    
    \begin{subfigure}[b]{0.47\textwidth}
        \centering
        \includegraphics[width=\textwidth]{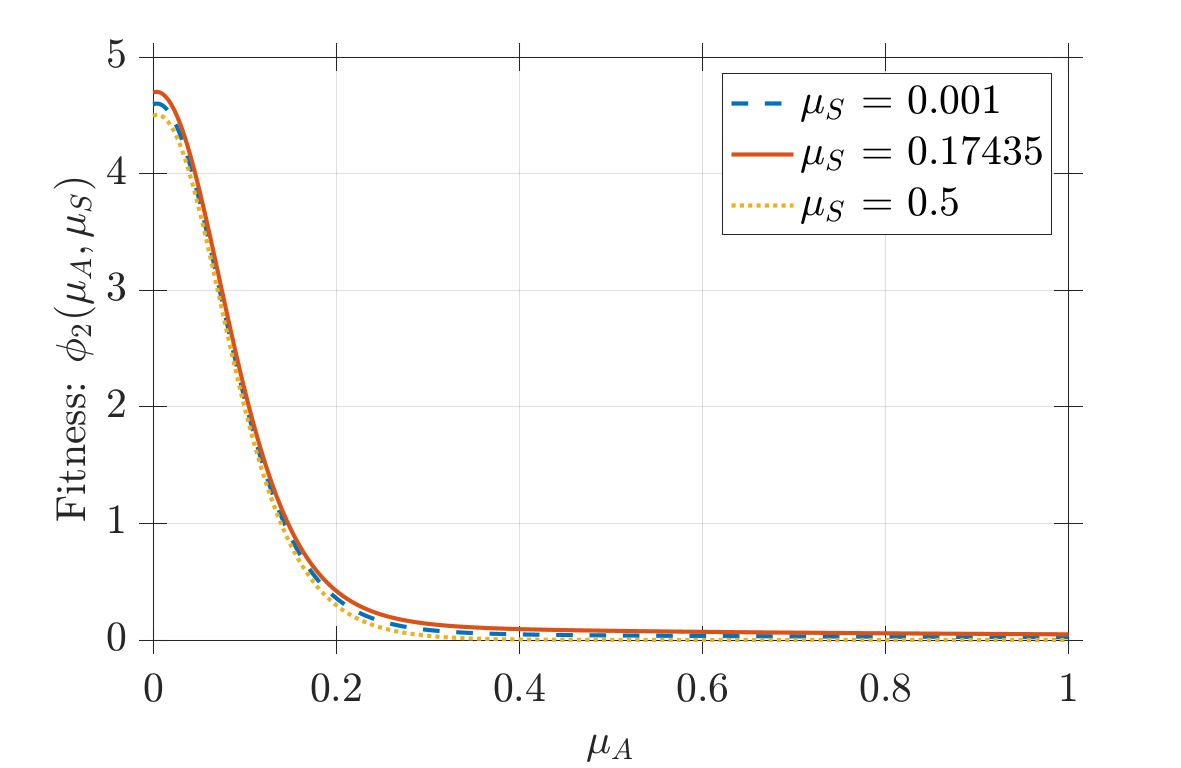}
        \caption{}\label{fig:1dfigmuAphi2}
    \end{subfigure}
    \begin{subfigure}[b]{0.47\textwidth}
        \centering
        \includegraphics[width=\textwidth]{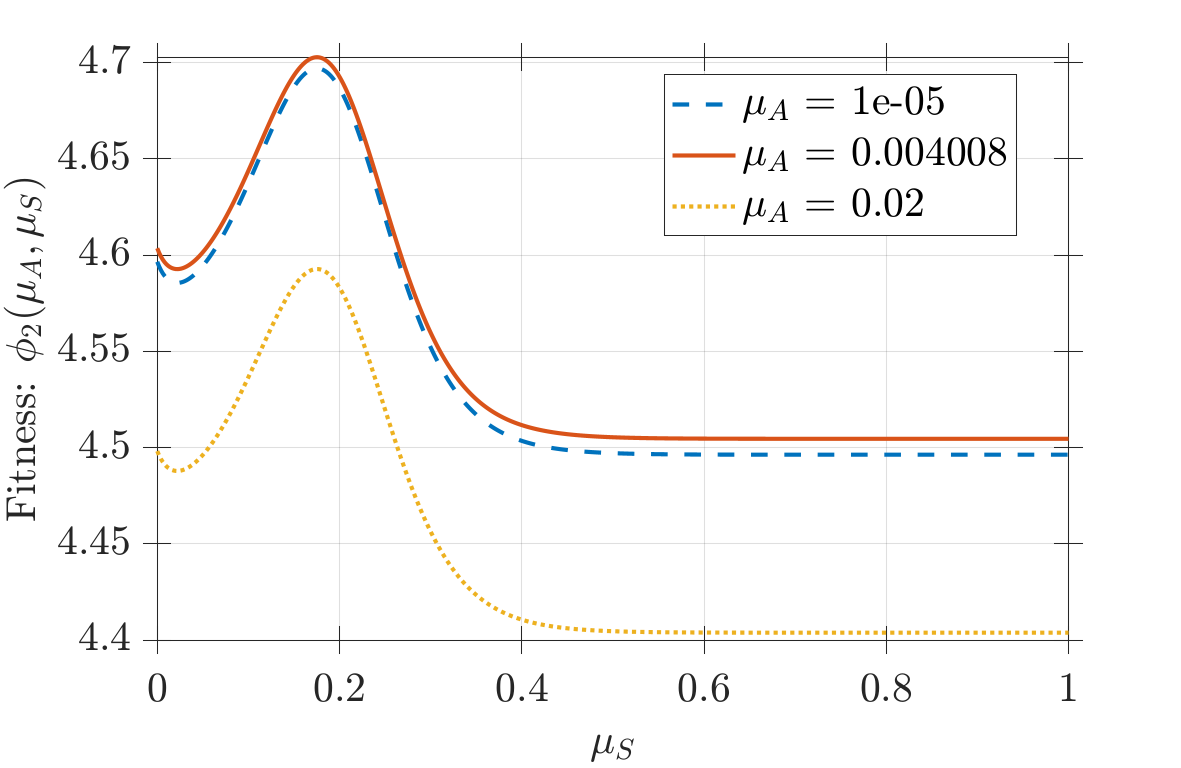}
        \caption{}\label{fig:1dfigmuSphi2}
    \end{subfigure}
    \caption{\textbf{Fitness function in the model with asymptomatic transmission.} Illustration of the fitness function $\phi_2$ for the model incorporating asymptomatic transmission: (a) The two-dimensional domain of the fitness function $\phi_2(\mu_A, \mu_S)$; (b) fitness function $\phi_2$ plotted with $\mu_A$ as the evolutionary variable, where each curve represents a different $\mu_S$ value, and the solid (red) line corresponds to the critical $\mu_S$ value that maximizes fitness; (c) fitness function $\phi_2$ plotted with $\mu_S$ as the evolutionary variable, where each curve represents a different $\mu_A$ value, and the solid (red) line corresponds to the critical $\mu_A$ value that maximizes fitness.
}\label{fig:fitness2}
\end{figure}

The fitness function for Model 1, using the specified parameters and the transmission function $\beta$, is illustrated in \cref{fig:ftness1}. The \gls{ess} for this model is achieved at the maximum value of the fitness function. As shown in \cref{fig:ftness1}, the maximum is attained at $\mu_1^* = 0.17435$, with a corresponding maximum fitness value of $4.5629$. Therefore, under the given parameter configuration the \gls{ess} is determined to be $\mu_1^* = 0.17435$. When extending the model to incorporate asymptomatic transmission, we introduce an additional evolutionary variable, $\mu_A$, along with the symptomatic virulence, $\mu_S$. As a result, the domain of the fitness function $\phi_2$ for this model becomes two-dimensional (see \cref{fig:2dfigphi2}). To further illustrate the behavior of the function, \cref{fig:1dfigmuAphi2} depicts the fitness function $\phi_2$ with $\mu_A$ as the evolutionary variable, where each curve in the plot corresponds to a different $\mu_S$ value. The curve represented by the solid (red) line corresponds to the critical $\mu_S$ value that achieves the maximum fitness. Similarly, \cref{fig:1dfigmuSphi2} shows the fitness function $\phi_2$ with $\mu_S$ as the evolutionary variable, where each curve in the plot corresponds to a different $\mu_A$ value. The curve represented by the solid (red) line corresponds to the critical $\mu_A$ value that achieves the maximum fitness.

As with Model 1, the \gls{ess} for this model is given by the point $(\mu_A^*, \mu_S^*)$ that maximizes the fitness function $\phi_2$. Using the chosen parameter values and transmission functions, the virulence values at the \gls{ess} are $\mu_A^* = 0.004008$ and $\mu_S^* = 0.17435$, with the maximum fitness value $\phi_2(\mu_A^*, \mu_S^*) = 4.7028$. Since the transmission function $\beta$ is identical to that of Model 1, the \gls{ess} value of the symptomatic virulence in Model 2 equals the virulence of the infectious population in Model 1 ($\mu_S^* = \mu_1^*$). However, in this case, the incorporation of the asymptomatic virulence, $\mu_A$, as an evolutionary variable allows for an increase in the fitness value, $\phi_2$, of the pathogen compared to the fitness value, $\phi_1$, in the simpler SEIR model.

\begin{figure}[htbp]
    \centering
    \begin{subfigure}[b]{0.45\textwidth}
        \centering
        \includegraphics[width=\textwidth]{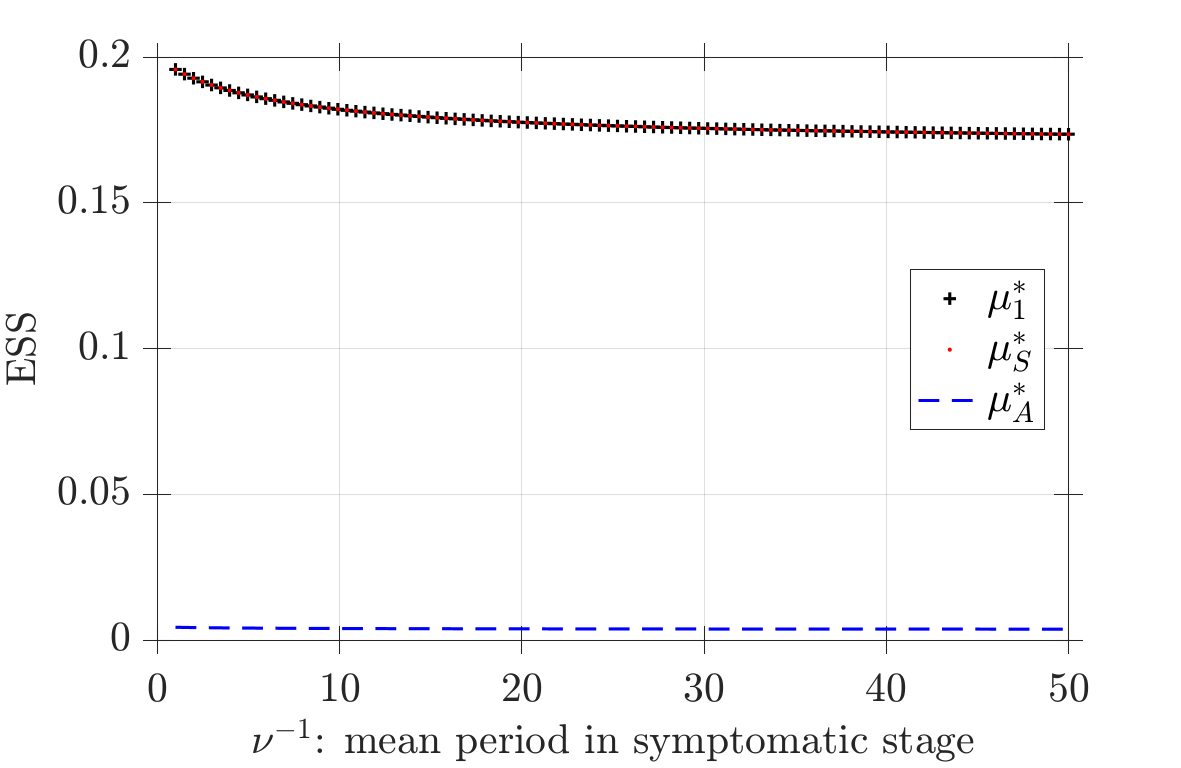}
        \caption{\gls{ess} for both models with $\nu^{-1}$}\label{fig:nu2M}
    \end{subfigure}
        \begin{subfigure}[b]{0.45\textwidth}
        \centering
        \includegraphics[width=\textwidth]{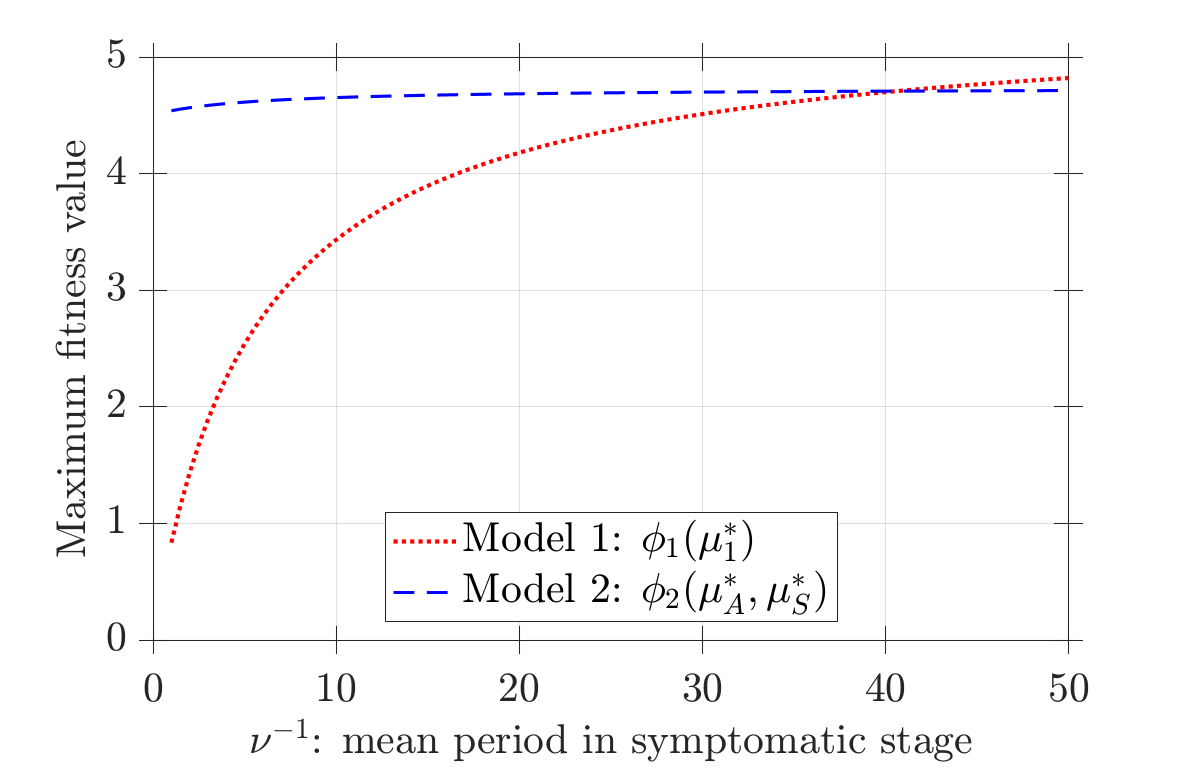}
        \caption{Maximum fitness values for both models with $\nu^{-1}$}\label{fig:nu2Fit}
    \end{subfigure}
    
    \begin{subfigure}[b]{0.32\textwidth}
        \centering
        \includegraphics[width=\textwidth]{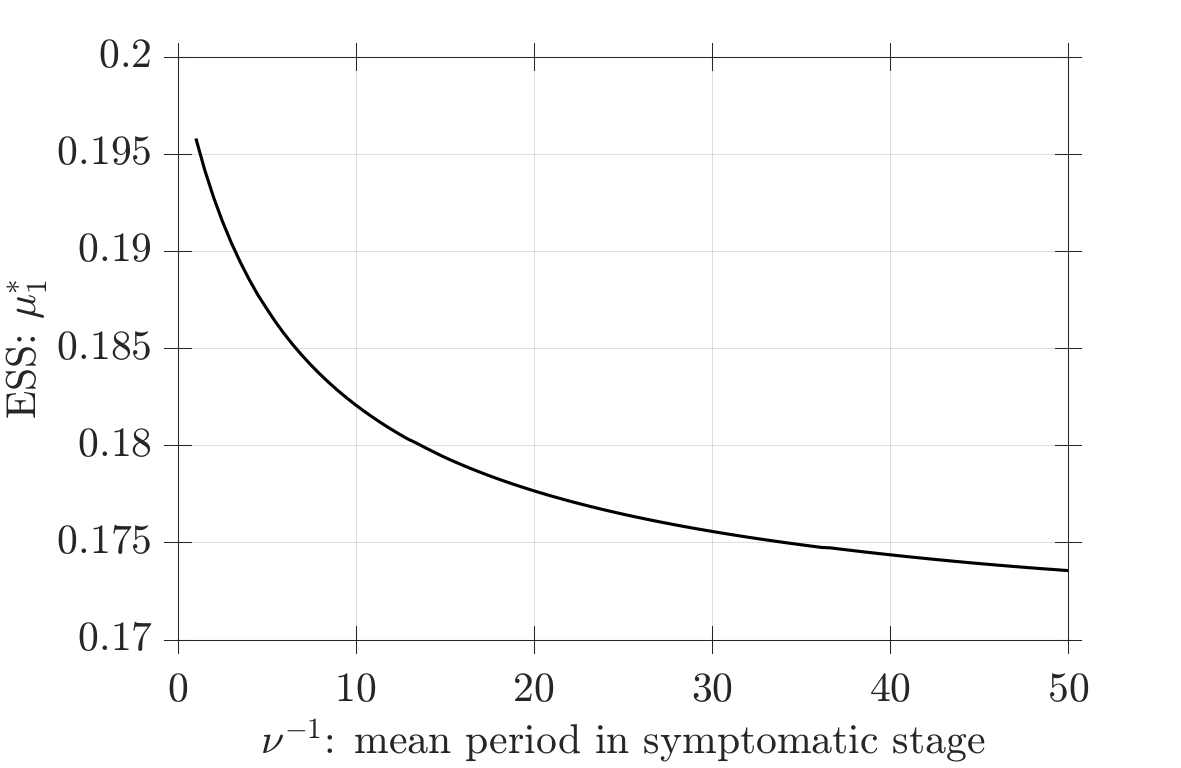}
        \caption{\gls{ess} $\mu_1^*$ for model 1 with $\nu^{-1}$}\label{fig:nu1M1}
    \end{subfigure}
        \begin{subfigure}[b]{0.32\textwidth}
        \centering
        \includegraphics[width=\textwidth]{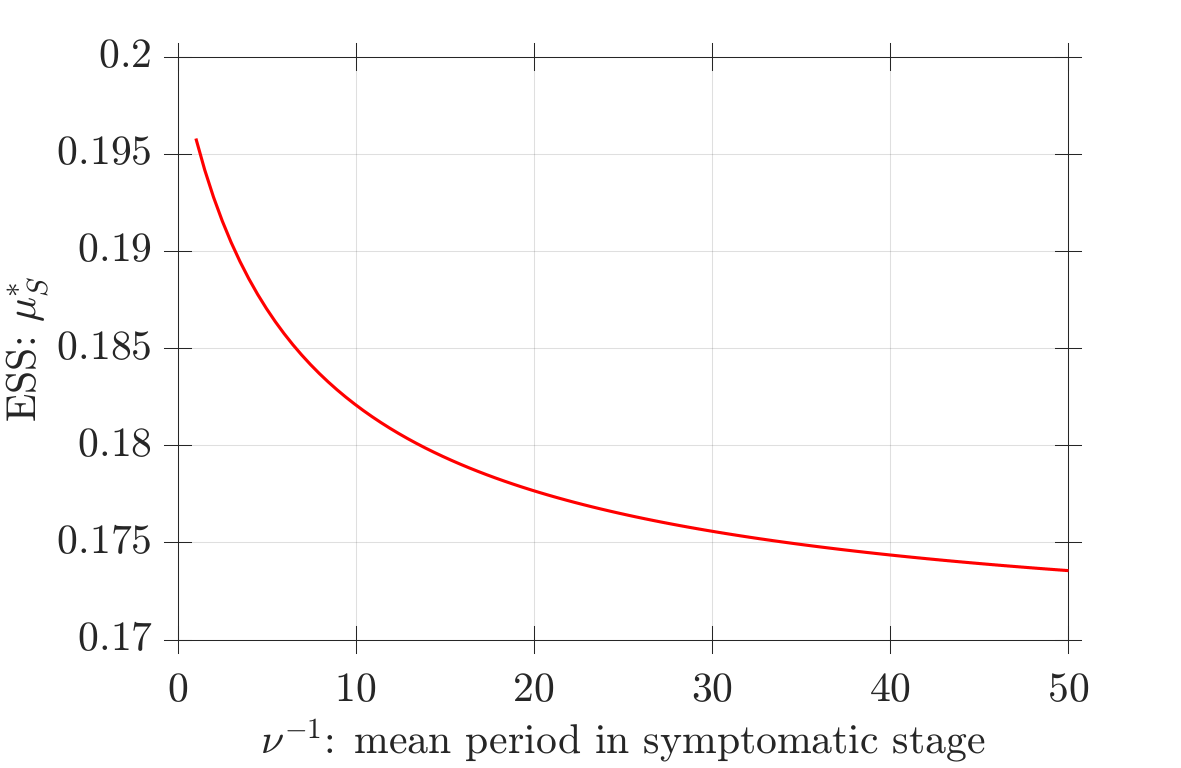}
        \caption{\gls{ess} $\mu_S^*$ for model 2 with $\nu^{-1}$}\label{fig:nu1M2MuS}
    \end{subfigure}
    \begin{subfigure}[b]{0.32\textwidth}
        \centering
        \includegraphics[width=\textwidth]{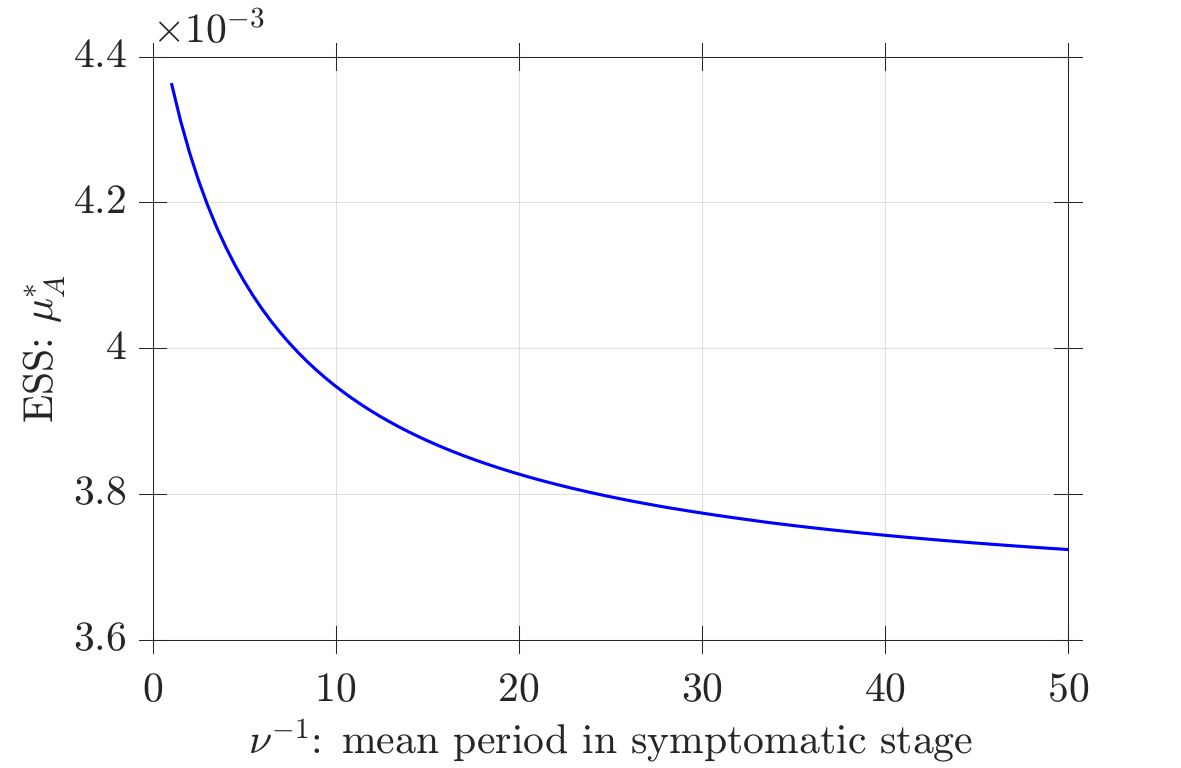}
        \caption{\gls{ess} $\mu_A^*$ for model 2 with $\nu^{-1}$}\label{fig:nu1M2MuA}
    \end{subfigure}
    \caption{\textbf{Impact of mean infectious period.} Comparison of the effects of the mean infectious period ($\nu^{-1}$) on the \gls{ess} and maximum fitness values for both models. (a) \gls{ess} values $\mu_1^*$ (model 1) and $\mu_A^*$, $\mu_S^*$ (model 2) as a function of $\nu^{-1}$. The curves for $\mu_1^*$ and $\mu_S^*$ are identical due to the same transmission function $\beta$ in both cases. The dashed (blue) curve represents $\mu_A^*$, which is significantly lower compared to $\mu_S^*$ or $\mu_1^*$ for corresponding $\nu^{-1}$ values. (b) Maximum fitness values for both models, showing greater stability for model 2. Model 1 fitness starts lower, increases rapidly, and surpasses model 2 around $\nu^{-1} \approx 40$. (c) and (d) Detailed views of \gls{ess} values $\mu_1^*$ and $\mu_S^*$, respectively, with respect to $\nu^{-1}$. (e) Decreasing relationship between $\mu_A^*$ and $\nu^{-1}$ for model 2, emphasizing its distinct evolutionary behavior.
}\label{fig:SenNu}
\end{figure}

Next, we focus on the effects of parameters on the \gls{ess} of both models. Specifically, we numerically investigate the impact of the mean infectious period ($\nu^{-1}$) in model 1 or the symptomatic infectious period in model 2 on the \gls{ess} for both models. Notably, this parameter is common to both models. As expected from the analytical discussion provided in the previous section, the \gls{ess} decreases as ($\nu^{-1}$) increases (see \cref{fig:nu2M,fig:nu1M1,fig:nu1M2MuS,fig:nu1M2MuA}). Furthermore, \cref{fig:nu2M} compares the behavior of the \gls{ess} values $\mu_1^*$ for model 1 and $\mu_A^*, \mu_S^*$ for model 2 as a function of $\nu^{-1}$. Since the same transmission function $\beta$ is chosen for both model 1 and the symptomatic group in model 2, the \gls{ess} curves for $\mu_1^*$ and $\mu_S^*$ are identical. A more detailed view of these curves is presented in \cref{fig:nu1M1,fig:nu1M2MuS}, respectively. In contrast, the \gls{ess} value of $\mu_A^*$ in model 2 is significantly lower compared to $\mu_S^*$ or $\mu_1^*$ for the corresponding $\nu^{-1}$ value (see the dashed (blue) curve in \cref{fig:nu2M}). A clear decreasing relationship with $\nu^{-1}$ is depicted in \cref{fig:nu1M2MuA}. 

We examine the maximum fitness values for both models in \cref{fig:nu2Fit}. This figure demonstrates that the maximum fitness value of model 2 is much more stable compared to model 1. For model 1, the maximum fitness value is initially lower than that of model 2 when the mean infectious period is short. However, it increases rapidly at the beginning and slows as the infectious period becomes longer. With the chosen parameter values, the maximum fitness value of model 1 surpasses that of model 2 around $\nu^{-1} \approx 40$. This observation highlights two key features: first, the maximum fitness for model 2 is more stable with respect to $\nu^{-1}$, and second, when the mean infectious period is sufficiently long, model 1 exhibits a higher maximum fitness compared to model 2, indicating that the influence of asymptomatic transmission diminishes in such scenarios.

Finally, we focus on the parameters specific to the asymptomatic model. We numerically analyze the effects of $\omega^{-1}$, the expected duration in the asymptomatic state, and $p$, the fraction of individuals following the ``mild" recovery pathway, on the evolutionary variables $\mu_A^*$ and $\mu_S^*$ at their stable strategies, as well as on the maximum fitness values for both models. It is important to note that the fitness value of model 1 remains unaffected by these parameters. However, we include model 1 in our analysis to facilitate a comparative evaluation of the fitness variations in model 2.

\begin{figure}[htbp]
    \centering
    \begin{subfigure}[b]{0.32\textwidth}
        \centering
        \includegraphics[width=\textwidth]{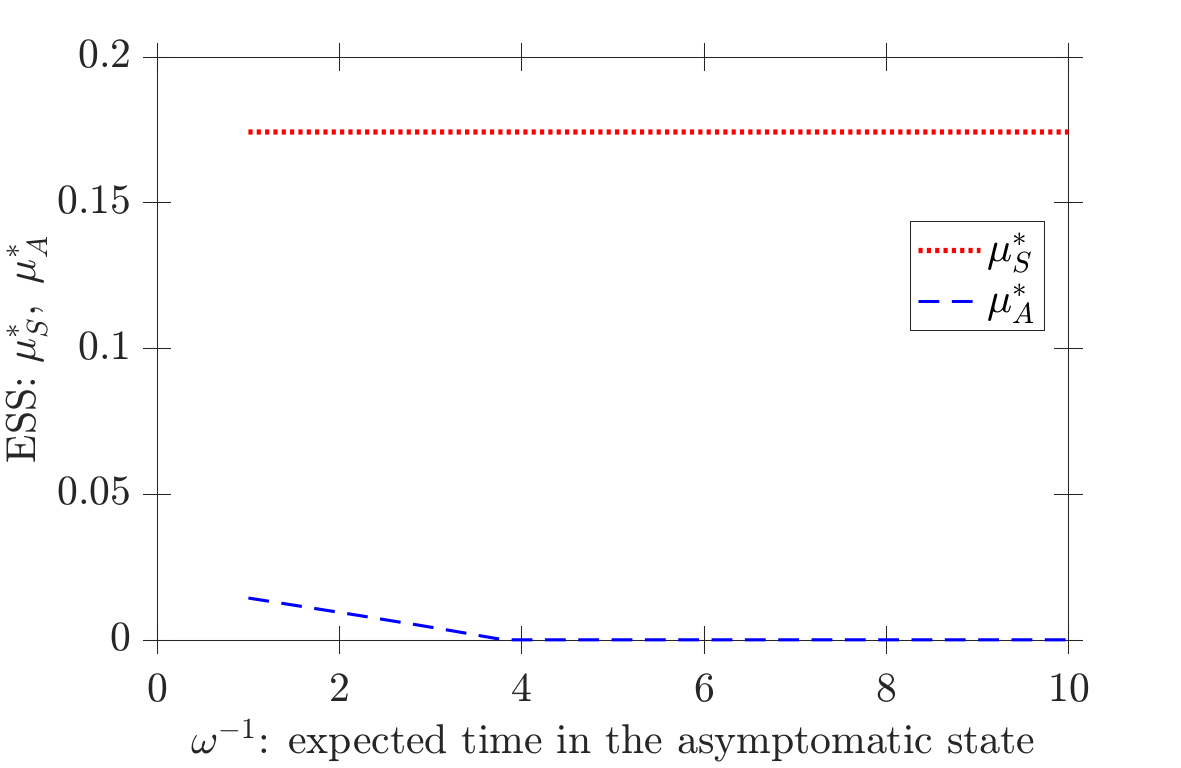}
        \caption{\gls{ess} with $\omega^{-1}$}\label{fig:omg1}
    \end{subfigure}
        \begin{subfigure}[b]{0.32\textwidth}
        \centering
        \includegraphics[width=\textwidth]{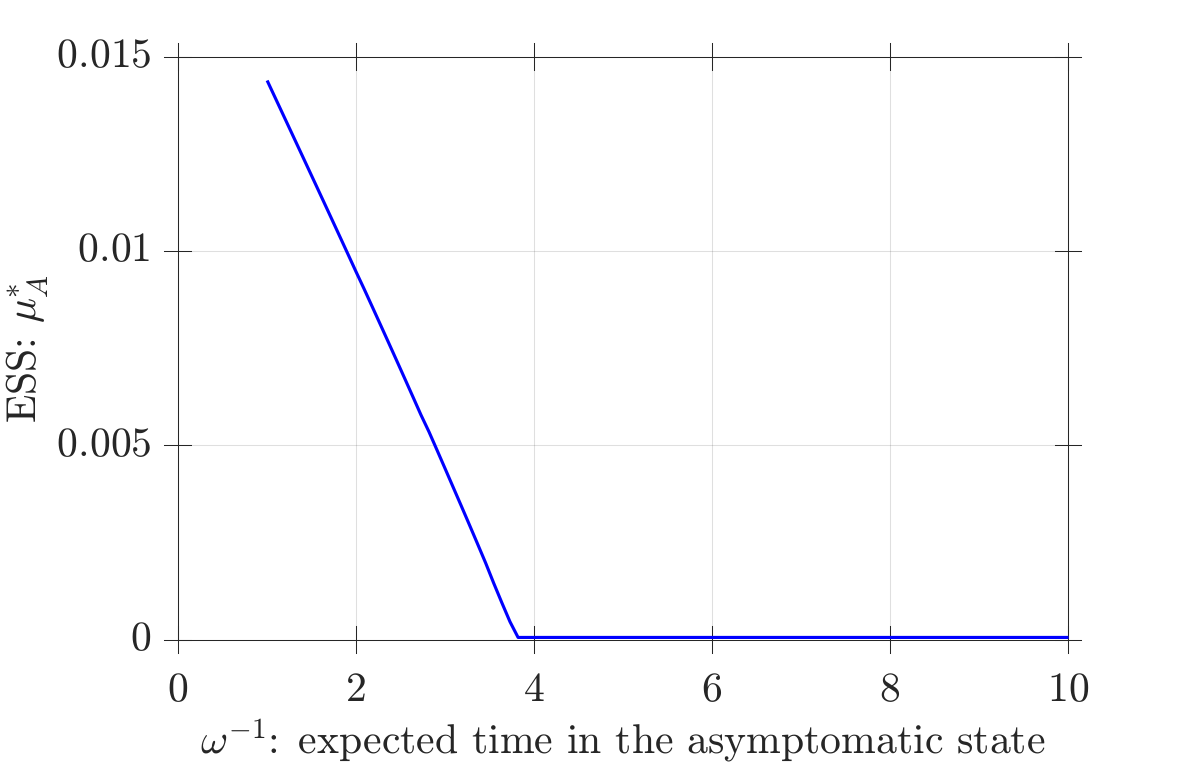}
        \caption{$\mu_A^*$ with $\omega^{-1}$}\label{fig:omg2}
    \end{subfigure}
            \begin{subfigure}[b]{0.32\textwidth}
        \centering
        \includegraphics[width=\textwidth]{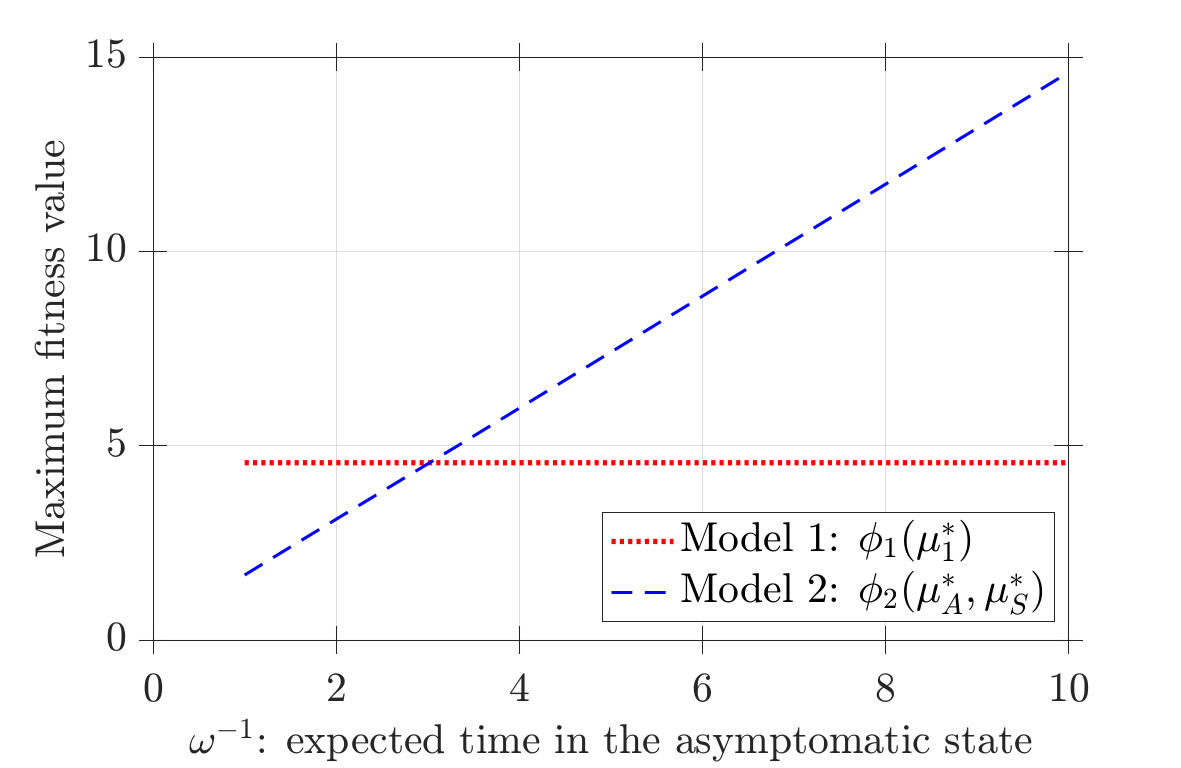}
        \caption{Maximum fitness with $\omega^{-1}$}\label{fig:omg3}
    \end{subfigure}
    
    \begin{subfigure}[b]{0.32\textwidth}
        \centering
        \includegraphics[width=\textwidth]{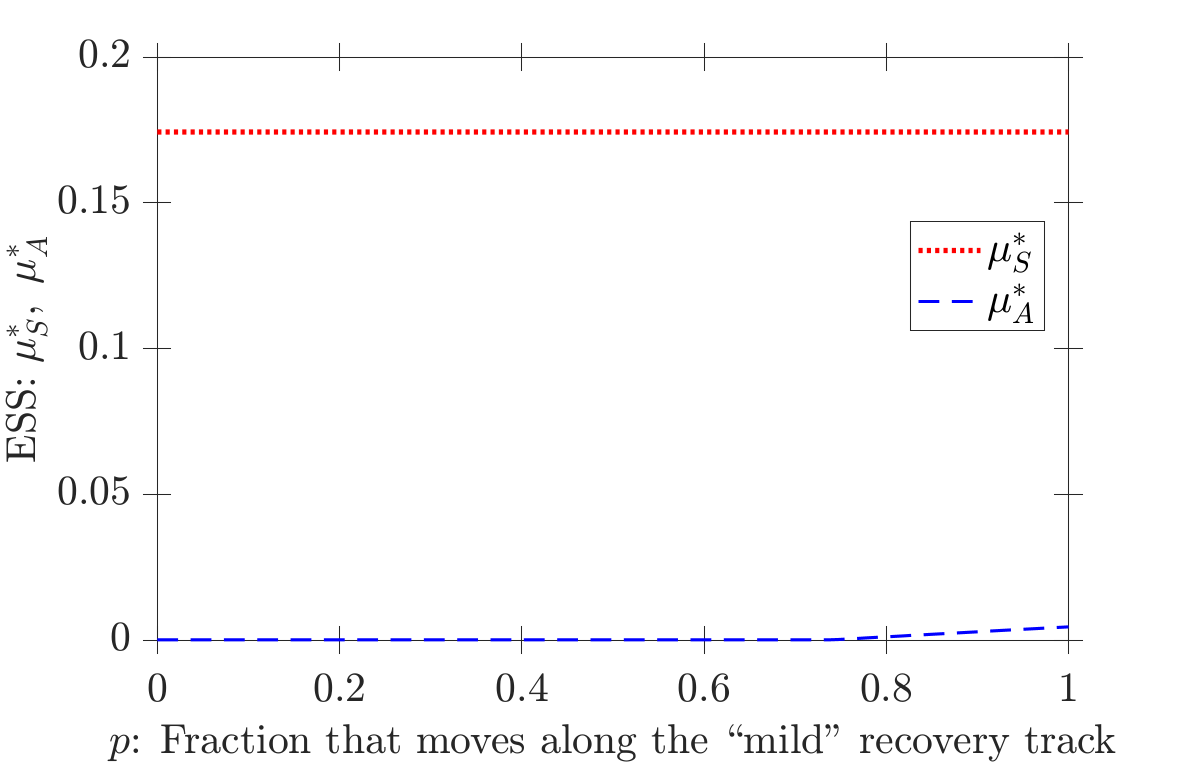}
        \caption{\gls{ess} with $p$}\label{fig:p1}
    \end{subfigure}
        \begin{subfigure}[b]{0.32\textwidth}
        \centering
        \includegraphics[width=\textwidth]{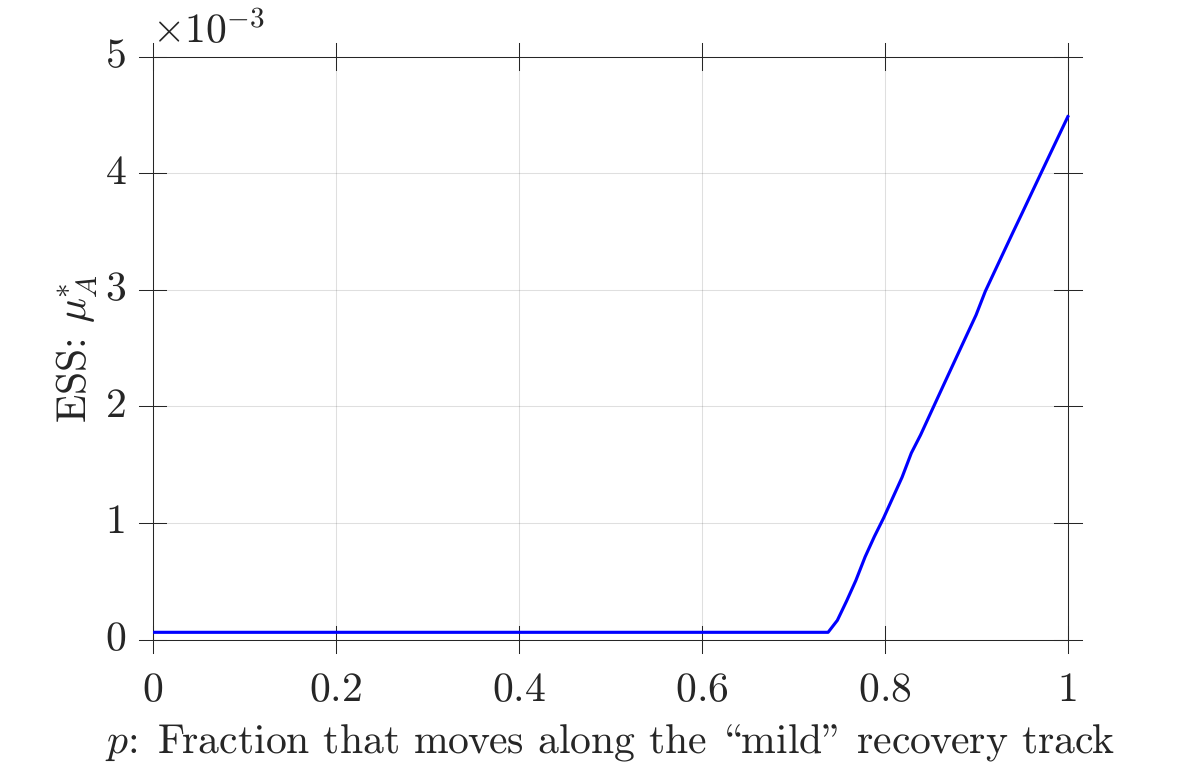}
        \caption{$\mu_A^*$ with $p$}\label{fig:p2}
    \end{subfigure}
    \begin{subfigure}[b]{0.32\textwidth}
        \centering
        \includegraphics[width=\textwidth]{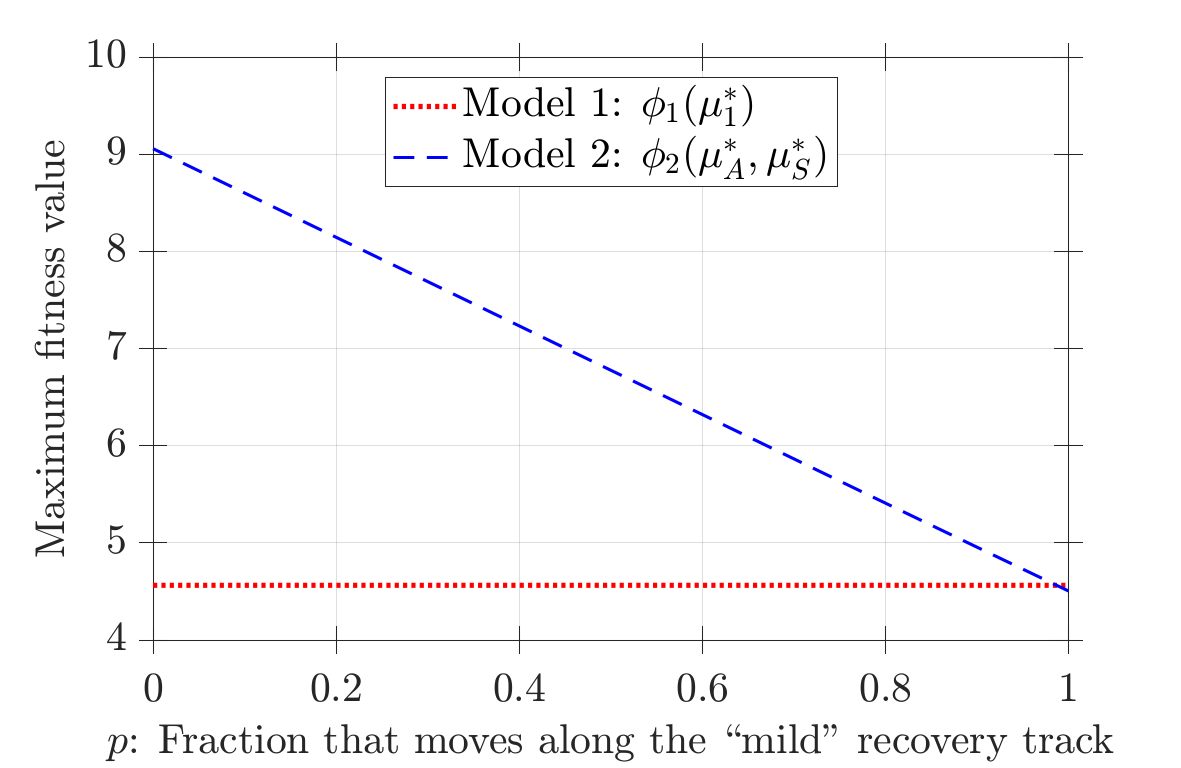}
        \caption{Maximum fitness with $p$}\label{fig:p3}
    \end{subfigure}
    \caption{\textbf{Impact of asymptomatic duration and mild recovery fraction.} The effect of the expected time in the asymptomatic state ($\omega^{-1}$) and the fraction of individuals following the ``mild" recovery pathway ($p$) on the \gls{ess} values and maximum fitness. (a) The \gls{ess} values $\mu_A^*$ and $\mu_S^*$ as functions of $\omega^{-1}$. (b) Detailed behavior of $\mu_A^*$ as a function of $\omega^{-1}$. (c) Maximum fitness values for model 2 as $\omega^{-1}$ increases, surpassing the fitness value of model 1 at $\omega^{-1} = 3$. (d) The \gls{ess} values $\mu_A^*$ and $\mu_S^*$ as functions of $p$. The \gls{ess} value $\mu_A^*$ remains significantly lower than $\mu_S^*$, with a slight increase for higher values of $p$. (e) Detailed behavior of $\mu_A^*$ with respect to $p$. (f) Maximum fitness values for model 2 decrease with increasing $p$ and converge to the maximum fitness value of model 1 as $p$ approaches 1. These results highlight the evolutionary dynamics favoring prolonged asymptomatic states and transitions to symptomatic states over direct recovery for maximizing pathogen fitness.}\label{fig:OmgP}
\end{figure}

\Cref{fig:omg1,fig:omg2,fig:omg3} illustrate the effect of $\omega^{-1}$, the expected duration in the asymptomatic state, while \Cref{fig:p1,fig:p2,fig:p3} show the effect of $p$, the fraction of individuals following the ``mild" recovery pathway, on the \gls{ess} values and the maximum fitness. The \gls{ess} value of $\mu_S^*$ for the symptomatic group in model 2 is equivalent to the \gls{ess} value $\mu_1^*$ in model 1 and remains unchanged with variations in either $\omega^{-1}$ (\Cref{fig:omg1}, dotted (red) curve) or $p$ (\Cref{fig:p1}, dotted (red) curve). Conversely, the \gls{ess} value of $\mu_A^*$, representing the asymptomatic group, is significantly smaller compared to $\mu_S^*$ across all parameter variations (see \Cref{fig:omg1,fig:p1}).

The relationship between the \gls{ess} value $\mu_A^*$ and the expected time in the asymptomatic state is depicted in \Cref{fig:omg2}, with the corresponding maximum fitness values shown in \Cref{fig:omg3}. Notably, as the expected time spent in the asymptomatic state increases ($\omega^{-1}$), the \gls{ess} value of $\mu_A^*$ decreases and approaches zero (\Cref{fig:omg2}. However, the corresponding maximum fitness value increases (\Cref{fig:omg3}) as $\omega^{-1}$ increases. Furthermore, with the selected parameter configuration, the maximum fitness value of model 2 surpasses that of model 1 around $\omega^{-1} = 3$ (\Cref{fig:omg3}), which aligns with the estimated parameter values in \cite{ogbunugafor2020variation}.

As shown in \Cref{fig:p3}, with the selected parameter values, an increase in $p$ results in the \gls{ess} value of $\mu_A^*$ remaining mostly unchanged or showing a very slight increase until a critical $p$ value is reached, after which it increases more significantly. However, the value of \gls{ess} $\mu_A^*$ remains much smaller compared to $\mu_S^*$ for any value of $p$. Interestingly, the maximum fitness value for model 2 (which contains both asymptomatic and symptomatic transmission routes) (\Cref{fig:p3}) decreases as $p$ increases and becomes equal to the maximum fitness value of model 1 (no asymptomatic compartment) when $p$ approaches $1$. This observation suggests an evolutionary preference for transitioning through the symptomatic transmission pathway rather than directly recovering after the asymptomatic state.

Based on these numerical investigations of $\omega^{-1}$ and $p$ with the selected parameter configuration, we can draw the following conclusions regarding the amount of time spent in the asymptomatic state. Intuitively, the transition from the asymptomatic state to the symptomatic state provides ($0 \lesssim p$) a higher pathogen fitness than directly ($p \lesssim 1$ ) moving from the asymptomatic to recovery (mild recovery track) (see \cref{fig:p3}). However, longer duration in the asymptomatic state results in increased pathogen fitness (see \cref{fig:omg3}), although with lower virulence (see \cref{fig:omg1}). Thus, in multiple scenarios, selection for prolonged time in the asymptomatic state facilitates higher fitness.

\color{black} 

\section{Discussion}
In this study, we used evolutionary invasion analysis to ask questions about optimal virulence levels in mathematical models of an infectious disease characterized by an asymptomatic transmission phase. We focus on virulence, which we define as the pathogen traits which make a person sick enough to be taken out of population circulation, such as being bedridden and physically ill. We examine several scenarios with real world relevance: when hosts have different mean infectious periods (due to some being asymptomatic carriers), or when public health interventions influence the time spent in the symptomatic or asymptomatic state.

Our analysis provides several notable points about optimal virulence strategies. We provide a mathematical basis for the (otherwise intuitive) evolutionary advantages of asymptomatic transmission: In settings where the asymptomatic window is prolonged, the fitness of the pathogen is higher (see \cref{fig:omg3}) when hosts progress through the symptomatic stage. However, in a scenario where an increased proportion of cases go through a strictly asymptomatic transmission route (the mild infectious route)---as in the case where symptomatic individuals are removed from the dynamics (via social distancing or treatment), or when features of hosts render them less likely to progress toward symptoms, optimal evolved strategies changes present a new scenario: the model structure rewires into one where asymptomatic transmission becomes the main (or exclusive) route (for details, refer to \cref{sec:ComModel}). However, the virulence of this new asymptomatic compartment (conferred by a mutant strain) is higher than the pre-evolved asymptomatic virulence level (see \cref{fig:p2}), though far lower than the symptomatic virulence level (see \cref{fig:p1}). The implications are that public health interventions can select for a different natural history of disease, where asymptomatic transmission becomes predominant, but where symptoms may slightly worsen (but again, less than the symptomatic state of the symptomatic class). The results highlight the interaction between public health intervention and pathogen evolution. Our attempts to mitigate symptoms and decrease symptomatic transmission events can lead to a less severe disease natural history. This prediction is highly contingent on model parameters and context-dependent features of disease model, but is a result to keep in mind as we continue to explore how different routes of infection drive different evolutionary optima for pathogens. 

We also observe that in models with asymptomatic transmission, an increase in time spent in the symptomatic window (the mean infectious period, $v^{-1}$) has a negligible effect on viral fitness (see \cref{fig:nu2Fit}), because transmission occurs through the asymptomatic compartment. This suggests that the asymptomatic route relaxes the selection of pathogen traits that influence the symptomatic window, as it is no longer the primary target of selection for optimal fitness. In addition, this scenario also predicts a slight decrease in virulence for both the asymptomatic and symptomatic compartments in both models (see \cref{fig:nu2M,fig:nu1M1,fig:nu1M2MuS,fig:nu1M2MuA}). The physics of this finding is intuitive: the lower the virulence, the more individuals remain in both compartments and are capable of transmitting. However, we re-emphasize that increasing the mean infectious period has a very slight impact on viral fitness. 

\subsection{Ideas and speculation}
How do we interpret these findings in light of contemporary conversations in disease evolution and mathematical epidemiology? One of the main results shows that increased flow through the mild recovery route facilitates a model dominated by asymptomatic transmission. This suggests that human interventions can influence virus evolution sufficiently for a fundamentally different structure (from S-E-IA-IS-R $\to$ S-E-IA-R). We acknowledge that mathematical models are simplified abstractions of reality. So, we should not overinterpret this finding, but rather explore the overall message: public interventions have evolutionary implications. In this case, asymptomatic transmission can facilitate a novel evolved disease ecology. In this case, the asymptomatic symptoms worsen slightly, but not to the level of the typical symptomatic state. Reflexively, this sounds like a positive outcome: evolution leads to a natural history with a less severe disease. And we should keep in mind that symptoms are not only markers of a negative health outcome, but also an organizational marker for public health intervention. In this sense, symptoms allow for rapid detection and facilitate rapid public health interventions. Ironically, a disease ecology with fewer noticeable symptoms can undermine our ability to intervene and lead to poor health outcomes, a finding observed in previous studies \cite{park2023intermediate}.

The robustness of viral fitness to changes in the mean infectious period (see \cref{fig:nu2Fit}) suggests that a complex disease history does not always promote the evolution of more virulent pathogens. From a clinical perspective, experiencing worse symptoms for a longer period is a less desirable outcome (for individuals) and may not be the optimal long-term outcome for a pathogen that emerges with complex transmission dynamics. But this also highlights how asymptomatic transmission derails expectations of how symptoms should evolve. That is, asymptomatic transmission can minimize the importance of symptomatic transmission, undermining one of the bases of a trade-off between transmission and virulence. This is not unlike the case of fomite or vector-borne transmission, where increased transmission can evolve in the presence of high virulence \cite{ewald1991waterborne, gandon1998curse}. This is a deceptively important notion: decoupling transmission success from symptoms could mean that symptoms are not the product of direct manipulation by the pathogen, but could nonetheless arise from idiosyncratic features of the host-parasite interaction. 


\subsection{Limitations}
Compartment models risk oversimplifying deeply complex notions and dynamics \cite{melikechi2022limits, roberto2021sars}. However, they are a useful tool for examining features of disease dynamics based on a set of transparent assumptions (all of which can be articulated analytically) \cite{weiss2013sir, otto2011biologist}. Although the transparency offered by compartmental models is appropriate for the specific questions that we examined in this study, we acknowledge that a "many models" approach, in which multiple frameworks are used to study a single complex system, can help address flaws in any given model type \cite{page2018model}.  

Secondly, our modeling approaches (like many others) are engineered with assumptions that undermine the direct application of these findings to any specific epidemiological setting. They are, however, relevant for a theoretical understanding of how asymptomatic transmission can influence evolutionary trajectories. 

Lastly, this study is limited by the lack of clarity surrounding notions such as virulence and asymptomatic transmission. With regard to the former, the term can have different meanings in different host-parasite contexts \cite{kabengele2024meta, acevedo2019virulence, franz2025trade}.  With regard to the latter, the real-world meaning of asymptomatic is dubious and also system-dependent \cite{shaikh2023asymptomatic}. More generally, there is a very thin line between very few symptoms and enough symptoms to qualify as truly symptomatic. But these questions about terminology can become clearer from the examination of mechanistic models and can help refine ideas for future research \cite{page2018model}.


\subsection{Future directions}

The use of evolutionary stable strategies to model virulence has been criticized on the grounds that it only computes the optimal outcomes, rather than the dynamics en route to that optimum \cite{day2004general}. In evolutionary invasion analysis, it is generally assumed that the evolutionary process is slower than the epidemiological dynamics. In cases where mutations are rare, the rate of evolution is often much slower than the rate at which the disease spreads, which means that mutant strains do not rapidly invade the population unless they have a significant fitness advantage. However, if the ``evolutionary speed" term becomes too large, evolutionary dynamics may not reach equilibrium \cite{mcgill2007evolutionary}. When evolutionary processes occur too quickly, the pathogen can continuously evolve and adapt, preventing the disease from reaching a stable state and resulting in chaotic dynamics. Therefore, relaxing the assumption of a slow evolutionary process presents a promising avenue for future investigation. This is especially relevant for rapidly evolving pathogens, such as RNA viruses \cite{steinhauer_rapid_1987}, which can differ significantly in the rate of evolution as a function of size, genomic content, and other characteristics. \cite{duffy2008rates}. This direction may help answer questions related to how different pathogen rates of evolution interact with features of the host-parasite interaction. 

More generally, our findings urge more research into how model structures drive expectations for the outcome of pathogen evolution. Pathogen evolution has many constraints, and this study (focusing on asymptomatic transmission) and others highlight how the differences between indirect and direct transmission models drive different expectations for how we expect virulence to evolve \cite{surasinghe2024evolutionary}. In addition, analyses of the published literature have shown that the relationship between virulence and transmission is profoundly influenced by idiosyncratic aspects of host-parasite interaction\cite{kabengele2024meta, acevedo2019virulence, franz2025trade}, and how changes in model structure undermine long-term predictions of disease dynamics \cite{scarpino2019predictability}.

Finally, our findings support claims that mathematical models are useful tools to address fundamental questions in the study of infectious diseases \cite{lofgren_mathematical_2014}. In addition, they encourage future collaboration between empirical and computational approaches, as the former can add details at the host-parasite, cellular, and molecular scales. When combined with modeling approaches, we can more responsibly recapitulate the shape of disease dynamics and make more informed predictions that can aid decision support in public health. 





\section*{Data availability}
The data for the article can be found in the text, and the parameters developed from other studies are appropriately cited. The code for this work can be found at \url{https://github.com/OgPlexus/Asymptomatic1}.

\section*{Acknowledgments}
The authors would like to acknowledge J. Weitz, S. Scarpino, N. Grubaugh, E. Blackmore, S. Manivannan, and K. Kabengele for their helpful input on various aspects of the manuscript. This project was inspired by a close reading of the book \textit{Asymptomatic} (Johns Hopkins Press, 2024) by Joshua Weitz. For a review of the book, see Ogbunugafor, 2024 \cite{ogbunugafor2024modes}.

\section*{Funding}
The authors would like to acknowledge funding support from the The Mynoon Doro and Stephen Doro MD, PhD Family Private Foundation, and from the Robert Wood Johnson Foundation's "Ideas for an Equitable Future" program. 

\section*{Conflicts of interest}
The authors declare that they have no conflict of interest. 

\section*{Author contributions}
Project conception: C.B.O. Model development: C.B.O. and S.S. Analysis and interpretation: C.B.O. and S.S. Writing, first draft: C.B.O. and S.S. Writing, revisions: C.B.O. and S.S. Writing, first draft. C.B.O. and S.S.Supervision: C.B.O.

\clearpage
{\small \bibliography{refs}}

\end{document}

%% file: glossary.tex
\newabbreviation{ode}{ODE}{ordinary differential equation}

\newglossaryentry{ess}{
    name={ESS},
    description={evolutionary stable strategy},
    first={evolutionary stable strategy (ESS)},
    plural={ESSs},
    firstplural={evolutionary stable strategies (ESSs)},
}

\newabbreviation{dfe}{DFE}{disease-free equilibrium }
\newabbreviation{mfe}{MFE}{mutant-free equilibrium }
\newabbreviation{covid}{SARS-CoV-2}{severe acute respiratory syndrome coronavirus 2}
\newabbreviation{hcv}{HCV}{Hepatitis C virus}
\glsxtrnewsymbol[description={basic reproduction number}]{r0}{\ensuremath{\operatorname{\mathcal{R}_0}}}

%% file: main.bbl
\begin{thebibliography}{10}

\bibitem{vanderwaal2016heterogeneity}
VanderWaal KL and Ezenwa VO (2016).
\newblock Heterogeneity in pathogen transmission.
\newblock \emph{Functional Ecology} \textbf{30(10)}:1606--1622

\bibitem{lloyd2005superspreading}
Lloyd-Smith JO, Schreiber SJ, Kopp PE and Getz WM (2005).
\newblock Superspreading and the effect of individual variation on disease emergence.
\newblock \emph{Nature} \textbf{438(7066)}:355--359

\bibitem{althouse2020superspreading}
Althouse BM, Wenger EA, Miller JC, Scarpino SV, Allard A, H{\'e}bert-Dufresne L and Hu H (2020).
\newblock Superspreading events in the transmission dynamics of SARS-CoV-2: Opportunities for interventions and control.
\newblock \emph{PLoS biology} \textbf{18(11)}:e3000897

\bibitem{st2024nonlinear}
St-Onge G, H{\'e}bert-Dufresne L and Allard A (2024).
\newblock Nonlinear bias toward complex contagion in uncertain transmission settings.
\newblock \emph{Proceedings of the National Academy of Sciences} \textbf{121(1)}:e2312202121

\bibitem{kiss2006effect}
Kiss IZ, Green DM and Kao RR (2006).
\newblock The effect of contact heterogeneity and multiple routes of transmission on final epidemic size.
\newblock \emph{Mathematical biosciences} \textbf{203(1)}:124--136

\bibitem{meszaros_direct_2020}
Meszaros VA, Miller-Dickson MD, Junior FBA, Almagro-Moreno S and Ogbunugafor CB (2020).
\newblock Direct transmission via households informs models of disease and intervention dynamics in cholera.
\newblock \emph{PLOS ONE} \textbf{15(3)}:e0229837.
\newblock Publisher: Public Library of Science

\bibitem{WeitzAsymptomatic}
Weitz J (2024).
\newblock \emph{Asymptomatic; The Silent Spread of COVID-19 and the Future of Pandemics}.
\newblock Johns Hopkins Press, Baltimore, MD

\bibitem{han2021prevalence}
Han D and Li J (2021).
\newblock Prevalence of asymptomatic SARS-CoV-2 infection.
\newblock \emph{Annals of internal medicine} \textbf{174(2)}:284

\bibitem{oran2020prevalence}
Oran DP and Topol EJ (2020).
\newblock Prevalence of asymptomatic SARS-CoV-2 infection: a narrative review.
\newblock \emph{Annals of internal medicine} \textbf{173(5)}:362--367

\bibitem{park2020time}
Park SW, Cornforth DM, Dushoff J and Weitz JS (2020).
\newblock The time scale of asymptomatic transmission affects estimates of epidemic potential in the COVID-19 outbreak.
\newblock \emph{Epidemics} \textbf{31}:100392

\bibitem{harris2023time}
Harris JD, Park SW, Dushoff J and Weitz JS (2023).
\newblock How time-scale differences in asymptomatic and symptomatic transmission shape SARS-CoV-2 outbreak dynamics.
\newblock \emph{Epidemics} \textbf{42}:100664

\bibitem{althouse2015asymptomatic}
Althouse BM and Scarpino SV (2015).
\newblock Asymptomatic transmission and the resurgence of Bordetella pertussis.
\newblock \emph{BMC medicine} \textbf{13}:1--12

\bibitem{shaikh2023asymptomatic}
Shaikh N, Swali P and Houben RM (2023).
\newblock Asymptomatic but infectious--The silent driver of pathogen transmission. A pragmatic review.
\newblock \emph{Epidemics} \textbf{44}:100704

\bibitem{moghadas2017asymptomatic}
Moghadas SM, Shoukat A, Espindola AL, Pereira RS, Abdirizak F, Laskowski M, Viboud C and Chowell G (2017).
\newblock Asymptomatic transmission and the dynamics of Zika infection.
\newblock \emph{Scientific reports} \textbf{7(1)}:5829

\bibitem{jamrozik2019surveillance}
Jamrozik E and Selgelid MJ (2019).
\newblock Surveillance and control of asymptomatic carriers of drug-resistant bacteria.
\newblock \emph{Bioethics} \textbf{33(7)}:766--775

\bibitem{fost1992ethical}
Fost N (1992).
\newblock Ethical implications of screening asymptomatic individuals.
\newblock \emph{The FASEB journal} \textbf{6(10)}:2813--2817

\bibitem{saad2020dynamics}
Saad-Roy CM, Wingreen NS, Levin SA and Grenfell BT (2020).
\newblock Dynamics in a simple evolutionary-epidemiological model for the evolution of an initial asymptomatic infection stage.
\newblock \emph{Proceedings of the national academy of sciences} \textbf{117(21)}:11541--11550

\bibitem{saad2021evolution}
Saad-Roy CM, Grenfell BT, Levin SA, Van Den~Driessche P and Wingreen NS (2021).
\newblock Evolution of an asymptomatic first stage of infection in a heterogeneous population.
\newblock \emph{Journal of the Royal Society Interface} \textbf{18(179)}:20210175

\bibitem{ogbunugafor2024modes}
Ogbunugafor CB (2024).
\newblock On modes of disease transmission and the hidden shape of pandemics: A review of Asymptomatic by Joshua Weitz.
\newblock \emph{Virus Evolution} \textbf{10(1)}:veae109

\bibitem{bull_virulence_1994}
Bull JJ (1994).
\newblock {VIRULENCE}.
\newblock \emph{Evolution} \textbf{48(5)}:1423--1437

\bibitem{alizon_virulence_2009}
Alizon S, Hurford A, Mideo N and Van~Baalen M (2009).
\newblock Virulence evolution and the trade-off hypothesis: history, current state of affairs and the future.
\newblock \emph{Journal of Evolutionary Biology} \textbf{22(2)}:245--259

\bibitem{lipsitch_virulence_1997}
Lipsitch M and Moxon ER (1997).
\newblock Virulence and transmissibility of pathogens: what is the relationship?
\newblock \emph{Trends in Microbiology} \textbf{5(1)}:31--37

\bibitem{lenski_evolution_1994}
Lenski RE and May RM (1994).
\newblock The {Evolution} of {Virulence} in {Parasites} and {Pathogens}: {Reconciliation} {Between} {Two} {Competing} {Hypotheses}.
\newblock \emph{Journal of Theoretical Biology} \textbf{169(3)}:253--265

\bibitem{park2023intermediate}
Park SW, Dushoff J, Grenfell BT and Weitz JS (2023).
\newblock Intermediate levels of asymptomatic transmission can lead to the highest epidemic fatalities.
\newblock \emph{PNAS nexus} \textbf{2(4)}:pgad106

\bibitem{basu2009evolution}
Basu S and Galvani AP (2009).
\newblock The evolution of tuberculosis virulence.
\newblock \emph{Bulletin of Mathematical Biology} \textbf{71}:1073--1088

\bibitem{surasinghe2024evolutionary}
Surasinghe S, Kabengele K, Turner PE and Ogbunugafor CB (2024).
\newblock Evolutionary Invasion Analysis of Modern Epidemics Highlights the Context-Dependence of Virulence Evolution.
\newblock \emph{Bulletin of Mathematical Biology} \textbf{86(8)}:88

\bibitem{smith1973logic}
Smith JM and Price GR (1973).
\newblock The logic of animal conflict.
\newblock \emph{Nature} \textbf{246(5427)}:15--18

\bibitem{otto2011biologist}
Otto SP and Day T (2011).
\newblock A biologist's guide to mathematical modeling in ecology and evolution.
\newblock In \emph{A Biologist's guide to mathematical modeling in ecology and evolution}. Princeton University Press

\bibitem{RN5}
Otto SP and Day T (2007).
\newblock \emph{A biologist’s guide to mathematical modeling in ecology and evolution}.
\newblock Princeton University Press, Princeton

\bibitem{williams2021evolutionary}
Williams PD and Kamel SJ (2021).
\newblock Evolutionary invasion analysis in structured populations.
\newblock \emph{Evolutionary Biology} \textbf{48(4)}:422--427

\bibitem{dietz1993estimation}
Dietz K (1993).
\newblock The estimation of the basic reproduction number for infectious diseases.
\newblock \emph{Statistical methods in medical research} \textbf{2(1)}:23--41

\bibitem{macdonald1952analysis}
Macdonald G (1952).
\newblock The analysis of equilibrium in malaria.
\newblock \emph{Tropical diseases bulletin} \textbf{49(9)}:813--829

\bibitem{fine1993herd}
Fine PE (1993).
\newblock Herd immunity: history, theory, practice.
\newblock \emph{Epidemiologic reviews} \textbf{15(2)}:265--302

\bibitem{delamater2019complexity}
Delamater PL, Street EJ, Leslie TF, Yang YT and Jacobsen KH (2019).
\newblock Complexity of the basic reproduction number (R0).
\newblock \emph{Emerging infectious diseases} \textbf{25(1)}:1

\bibitem{van2008further}
Van~den Driessche P and Watmough J (2008).
\newblock Further notes on the basic reproduction number.
\newblock \emph{Mathematical epidemiology} pp. 159--178

\bibitem{boldin2012evolutionary}
Boldin B and Kisdi E (2012).
\newblock On the evolutionary dynamics of pathogens with direct and environmental transmission.
\newblock \emph{Evolution} \textbf{66(8)}:2514--2527

\bibitem{van1995dynamics}
van Baalen M and Sabelis MW (1995).
\newblock The dynamics of multiple infection and the evolution of virulence.
\newblock \emph{The American Naturalist} \textbf{146(6)}:881--910

\bibitem{diekmann1990definition}
Diekmann O, Heesterbeek JAP and Metz JA (1990).
\newblock On the definition and the computation of the basic reproduction ratio R 0 in models for infectious diseases in heterogeneous populations.
\newblock \emph{Journal of mathematical biology} \textbf{28}:365--382

\bibitem{diekmann2010construction}
Diekmann O, Heesterbeek J and Roberts MG (2010).
\newblock The construction of next-generation matrices for compartmental epidemic models.
\newblock \emph{Journal of the royal society interface} \textbf{7(47)}:873--885

\bibitem{castillo2020tour}
Castillo-Garsow CW and Castillo-Chavez C (2020).
\newblock A tour of the basic reproductive number and the next generation of researchers.
\newblock \emph{An Introduction to Undergraduate Research in Computational and Mathematical Biology: From Birdsongs to Viscosities} pp. 87--124

\bibitem{surasinghe2024structural}
Surasinghe S, Manivannan SN, Scarpino SV, Crawford L and Ogbunugafor CB (2024).
\newblock Structural causal influence (SCI) captures the forces of social inequality in models of disease dynamics.
\newblock \emph{arXiv preprint arXiv:240909096}

\bibitem{ogbunugafor2020variation}
Ogbunugafor CB, Miller-Dickson MD, Meszaros VA, Gomez LM, Murillo AL and Scarpino SV (2020).
\newblock Variation in microparasite free-living survival and indirect transmission can modulate the intensity of emerging outbreaks.
\newblock \emph{Scientific reports} \textbf{10(1)}:1--14

\bibitem{ewald1991waterborne}
Ewald PW (1991).
\newblock Waterborne transmission and the evolution of virulence among gastrointestinal bacteria.
\newblock \emph{Epidemiology \& Infection} \textbf{106(1)}:83--119

\bibitem{gandon1998curse}
Gandon S (1998).
\newblock The curse of the pharoah hypothesis.
\newblock \emph{Proceedings of the Royal Society of London Series B: Biological Sciences} \textbf{265(1405)}:1545--1552

\bibitem{melikechi2022limits}
Melikechi O, Young AL, Tang T, Bowman T, Dunson D and Johndrow J (2022).
\newblock Limits of epidemic prediction using SIR models.
\newblock \emph{Journal of Mathematical Biology} \textbf{85(4)}:36

\bibitem{roberto2021sars}
Roberto~Telles C, Lopes H and Franco D (2021).
\newblock SARS-COV-2: SIR model limitations and predictive constraints.
\newblock \emph{Symmetry} \textbf{13(4)}:676

\bibitem{weiss2013sir}
Weiss HH (2013).
\newblock The SIR model and the foundations of public health.
\newblock \emph{Materials matematics} pp. 0001--17

\bibitem{page2018model}
Page SE (2018).
\newblock \emph{The model thinker: What you need to know to make data work for you}.
\newblock Hachette UK

\bibitem{kabengele2024meta}
Kabengele K, Turner WC, Turner PE and Ogbunugafor CB (2024).
\newblock A meta-analysis highlights the idiosyncratic nature of tradeoffs in laboratory models of virus evolution.
\newblock \emph{Virus Evolution} \textbf{10(1)}:veae105

\bibitem{acevedo2019virulence}
Acevedo MA, Dillemuth FP, Flick AJ, Faldyn MJ and Elderd BD (2019).
\newblock Virulence-driven trade-offs in disease transmission: A meta-analysis.
\newblock \emph{Evolution} \textbf{73(4)}:636--647

\bibitem{franz2025trade}
Franz M, Armitage SA, McMahon D, Subasi BS and Rafaluk C (2025).
\newblock Trade-offs in virulence evolution: a Hierarchy-of-Hypotheses approach.
\newblock \emph{Trends in Parasitology}

\bibitem{day2004general}
Day T and Proulx SR (2004).
\newblock A general theory for the evolutionary dynamics of virulence.
\newblock \emph{The American Naturalist} \textbf{163(4)}:E40--E63

\bibitem{mcgill2007evolutionary}
McGill BJ and Brown JS (2007).
\newblock Evolutionary game theory and adaptive dynamics of continuous traits.
\newblock \emph{Annu Rev Ecol Evol Syst} \textbf{38(1)}:403--435

\bibitem{steinhauer_rapid_1987}
Steinhauer DA and Holland JJ (1987).
\newblock Rapid {Evolution} of {RNA} {Viruses}.
\newblock \emph{Annual Review of Microbiology} \textbf{41(1)}:409--431

\bibitem{duffy2008rates}
Duffy S, Shackelton LA and Holmes EC (2008).
\newblock Rates of evolutionary change in viruses: patterns and determinants.
\newblock \emph{Nature Reviews Genetics} \textbf{9(4)}:267--276

\bibitem{scarpino2019predictability}
Scarpino SV and Petri G (2019).
\newblock On the predictability of infectious disease outbreaks.
\newblock \emph{Nature communications} \textbf{10(1)}:898

\bibitem{lofgren_mathematical_2014}
Lofgren ET, Halloran ME, Rivers CM, Drake JM, Porco TC, Lewis B, Yang W, Vespignani A, Shaman J et~al. (2014).
\newblock Mathematical models: {A} key tool for outbreak response.
\newblock \emph{Proceedings of the National Academy of Sciences} \textbf{111(51)}:18095--18096.
\newblock Publisher: Proceedings of the National Academy of Sciences

\end{thebibliography}
